\documentclass[prb,twocolumn,amssymb,amsfonts,superscriptaddress,showpacs,floatfix]{revtex4}
\usepackage{epsfig}
\usepackage{amsmath}
\usepackage{bm}

\renewcommand{\theequation}{\thesection.\arabic{equation}}

\begin{document}
\title{Theory of neutral and charged exciton scattering with electrons in
semiconductor quantum wells}

\author{G.~Ramon}
\email{ramon@physics.technion.ac.il.}
\author{A.~Mann}
\affiliation{Department of Physics, Technion -- Israel Institute
of Technology, Haifa 32000, Israel}
\author{E.~Cohen}
\affiliation{Department of Physics, Technion -- Israel Institute
of Technology, Haifa 32000, Israel} \affiliation{Solid State
Institute, Technion -- Israel Institute of Technology, Haifa
32000, Israel}

\begin{abstract}

Electron scattering on both neutral ($X$) and charged ($X^-$)
excitons in quantum wells is studied theoretically. A microscopic
model is presented, taking into account both elastic and
dissociating scattering. The model is based on calculating the
exciton-electron direct and exchange interaction matrix elements,
from which we derive the exciton scattering rates. We find that
for an electron density of $10^9 {\rm cm}^{-2}$ in a GaAs QW at
$T=5K$, the $X^-$ linewidth due to electron scattering is roughly
twice as large as that of the neutral exciton. This reflects both
the $X^-$ larger interaction matrix elements compared with those
of $X$, and their different dependence on the transferred
momentum. Calculated reflection spectra can then be obtained by
considering the three electronic excitations of the system,
namely, the heavy-hole and light-hole 1S neutral excitons, and the
heavy-hole 1S charged exciton, with the appropriate oscillator
strengths.

\end{abstract}


\maketitle

\section{Introduction}
\label{intro}

The broadening of exciton emission lines due to electron-exciton
collisions was first observed in bulk GaAs \cite{Leit}. An
experimental comparison between exciton-electron and
exciton-exciton scattering mechanisms in bulk GaAs \cite{SchHon86}
and in GaAs quantum wells (QW's) \cite{SchHon89,Cap} showed that
the exciton-electron scattering efficiency is an order of
magnitude larger than that of the exciton-exciton process. Both
scattering processes are enhanced for the 2D excitons as compared
to bulk excitons. Theoretical calculations of elastic and
inelastic exciton-electron scattering were reported for bulk
semiconductors \cite{MatElk} and for QW's \cite{FenSpe}. The
semiclassical treatment of the latter neglected the exchange term
which was shown recently to be the dominant term in QW's
\cite{prb}.

The existence of the negatively charged exciton $X^-$ (trion) was
first proposed by Lampert \cite{Lamp} as the semiconductor
analogue of the hydrogen ion. Due to their confinement in the
growth direction, 2D trions have a binding energy which is an
order-of-magnitude larger than that of the bulk trions
\cite{Stebe1}. This fact facilitated the observation of trions in
CdTe \cite{Kheng} and in GaAs \cite{Fink} QW's containing a
low-density two-dimensional electron gas (2DEG). Theoretical
calculations of the binding energy of both negatively ($X^-$) and
positively ($X^+$) charged excitons were performed using various
trial functions, showing that only the singlet state is bound
\cite{Stebe1,Bob,Stebe}. Particular attention was devoted to the
modification of the trion's properties in the presence of strong
magnetic fields. In this limit, the $X^-$ binding energy is
increased, and the electron spin triplet state becomes bound as
well \cite{Chap}. It was also observed that the dependence on the
magnetic field of the $X^-$ and $X^+$ binding energies, as well as
their Zeeman splitting, differ drastically (while being nearly
identical at zero magnetic field) \cite{Glas}.

While these aspects of the trions where extensively studied, there
are very few reports on their broadening mechanisms
\cite{Amnon,Yayon}. In this paper we present a theoretical model
for the scattering of both neutral and charged excitons with
electrons in QW's. This paper continues our previous work which
introduced neutral excitons elastic scattering with electrons in
the context of cavity polaritons \cite{prb}. Here we extend our
model, to incorporate both excitons and trions, taking into
account elastic and inelastic scattering processes. The model is
based on calculating the exciton-electron direct and exchange
interaction matrix elements, from which we derive the exciton
scattering rates. These are integrated over all final excitonic
states, resulting in the exciton linewidth due to scattering, as a
function of its initial momentum (or energy). A major difference
between the exciton and trion scattering lies in the charge of the
latter, which results in a divergence of its matrix elements in
the limit of zero transferred momentum. This divergence,
originating from the infinite range of the Coulomb potential, is
treated by applying the Lindhard model for the potential
screening. The screening action complicates the trion linewidth
dependence on the electron density; thus, while for electron
densities larger than $5\cdot 10^9 {\rm cm}^{-2}$ the trion
linewidth due to electron scattering in a GaAs QW at $T=5K$ is
comparable to that of the neutral exciton, for a very dilute 2DEG
($n_{\rm e} \approx 5\cdot 10^8 {\rm cm}^{-2}$) the trion
linewidth becomes roughly an order-of-magnitude larger than that
of the neutral exciton. Inelastic scattering is also much more
efficient in the trion's case, due to its smaller binding energy.
The calculated reflection spectra are obtained by considering the
three electronic excitations of the system, namely, the heavy-hole
and light-hole 1S neutral excitons, and the heavy-hole 1S charged
exciton (trion), with the appropriate oscillator strengths. A
qualitative validation of our calculations is given by considering
photoluminesence (PL) measurements that were done on a mixed type
I -- type II GaAs/AlAs QW (MTQW) structure \cite{Amnon}.

The paper is organized as follows. In section \ref{ex-el}, we
present in detail the model for the neutral exciton - electron
scattering. In section \ref{trion} we consider the trion-electron
scattering. First we construct the trion's wave function, which is
used in the matrix elements calculations. We then treat the
divergences that originate from the infinite range of the Coulomb
potential, by applying the Lindhard model for the potential
screening. The effect of the screening is discussed in the context
of the trion linewidth dependence on the electron density. In
section \ref{bare-ex} we use our model to calculate QW reflection
spectra and discuss their relevance to the available experimental
data. A summary and conclusions are given in section \ref{conc}.
In Appendices A and B, we provide some details of the direct and
exchange integrals calculations for the neutral exciton - electron
scattering. Finally, a derivation of the trion binding energy,
using a Chandrasekhar-type trial function, is given in Appendix C.


\section{A microscopic model for exciton - electron scattering}
\label{ex-el}
\setcounter{equation}{0}

In this section we present a detailed description of the
exciton-electron scattering, considering separately the elastic
process, for which the exciton remains bound, and the inelastic
process, where the exciton breaks into an unbound electron-hole
pair.

\subsection{Elastic scattering}

\begin{figure}[htbp]
\epsfxsize=0.55\textwidth
\centerline{\epsffile{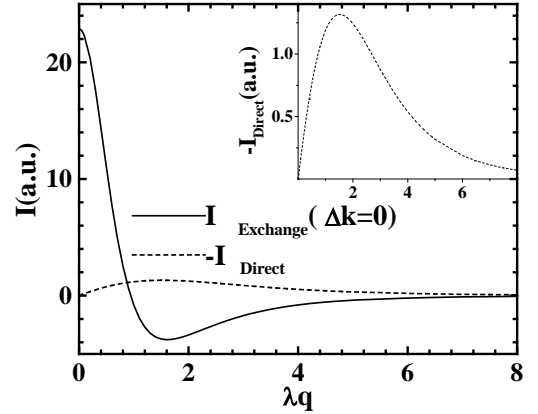}}
\vspace*{-4cm}
\caption{Calculated direct and exchange integrals vs. transferred
momentum. The inset shows $-I_{\rm dir}$ on an expanded scale.}
\label{dir_exc}
\end{figure}

We first construct the state of a single exciton with an in-plane
center-of-mass (CM) momentum ${\bf k}_{\rm x}$ in the fermionic
Hilbert space of electron-hole pairs, using the notations of
Tassone and Yamamoto \cite{TasYam99}. It is a superposition of
wave functions with different electron momenta ${\bf k}_1$ and
electron and hole $z$ coordinates, given by
\begin{equation}
| {\bf k}_{\rm x} \rangle = \sum_{{\bf k}_1} \int dz_e dz_h
\phi^*_{\alpha {\bf k}_{\rm x}+{\bf k}_1} (z_e,z_h) c^\dag_{-{\bf
k}_1,z_e} d^\dag_{{\bf k}_{\rm x}+{\bf k}_1,z_h} | 0 \rangle ,
\label{ex_st}
\end{equation}
where $c^\dag_{{\bf k}_{\rm x},z_e} \ (d^\dag_{{\bf k}_{\rm
x},z_h})$ is the electron (hole) creation operator with in-plane
momentum ${\bf k}_{\rm x}$ and $z_e$ ($z_h$) coordinate, $m_{\rm
e}$, $M_{\rm x}$ are the electron and exciton in-plane effective
masses, respectively, $\alpha=m_{\rm e}/M_{\rm x}$, and
$\phi_{{\bf k}} (z_e,z_h)={\cal N} \left[1+(\lambda k)^2
\right]^{-3/2}\chi_e(z_e) \chi_h(z_h)$ is the in-plane Fourier
transform of the exciton wave function (see Appendix A). A state
comprising an exciton and an unbound electron having ${\bf k}_{\rm
e}$ and $z_c$ will be written as
\begin{eqnarray}
| {\bf k}_{\rm x} ; {\bf k}_{\rm e} \rangle &=& \sum_{{\bf k}_1}
\int dz_e dz_h dz_c \phi^*_{\alpha {\bf k}_{\rm x}+ {\bf k}_1}
(z_e,z_h) \psi^*_{{\bf k}_{\rm e}} (z_c) \times \nonumber \\
&& c^\dag_{-{\bf k}_1,z_e} d^\dag_{{\bf k}_{\rm x}+{\bf k}_1,z_h}
c^\dag_{{\bf k}_{\rm e},z_c} | 0 \rangle ,
\label{initial}
\end{eqnarray}
where $\psi_{{\bf k}_{\rm e}} (z_c)$ is the electron wave
function. Applying the Coulomb interaction operators $V_{ee}$,
$V_{eh}$ to this state, we have for the electron-hole and
electron-electron interactions:
\begin{widetext}
\begin{subequations}
\label{VehVee}
\begin{eqnarray}
V_{eh} |{\bf k}_{\rm x} ; {\bf k}_{\rm e} \rangle &=& \sum_{{\bf
k}_1,{\bf q}_1} \int dz_e dz_h dz_c V_{{\bf q}_1}(z_h-z_c) \left\{
\phi^*_{\alpha {\bf k}_{\rm x}+ {\bf k}_1+{\bf q}_1}(z_e,z_h)
\psi^*_{{\bf k}_{\rm e}} (z_c) c^\dag_{-{\bf k}_1,z_e}
d^\dag_{{\bf k}_{\rm x}+{\bf k}_1,z_h} c^\dag_{{\bf k}_{\rm
e},z_c}
\right. \nonumber \\
&+& \left. \phi^*_{\alpha {\bf k}_{\rm x}+ {\bf k}_1}(z_e,z_h)
\psi^*_{{\bf k}_{\rm e}}(z_c) c^\dag_{-{\bf k}_1,z_e} d^\dag_{{\bf
k}_{\rm x}+{\bf k}_1-{\bf q}_1,z_h} c^\dag_{{\bf k}_{\rm e}+ {\bf
q}_1,z_c} \right\} |0 \rangle . \label{Veh} \\
V_{ee} |{\bf k}_{\rm x} ; {\bf k}_{\rm e} \rangle &=& \sum_{{\bf
k}_1,{\bf q}_1} \int dz_e dz_h dz_c V_{{\bf q}_1}(z_e-z_c)
\phi^*_{\alpha {\bf k}_{\rm x}+ {\bf k}_1}(z_e,z_h) \psi^*_{{\bf
k}_{\rm e}} (z_c) c^\dag_{-{\bf k}_1-{\bf q}_1,z_e} d^\dag_{{\bf
k}_{\rm x}+{\bf k}_1,z_h} c^\dag_{{\bf k}_{\rm e}+{\bf q}_1,z_c}
|0 \rangle .
\label{Vee}
\end{eqnarray}
\end{subequations}
The first term in Eq.\ (\ref{Veh}) represents the Coulomb
interaction between the constituents of the exciton, thus
contributing to its self energy, and can be discarded in the
calculation of the scattering matrix elements. In Eqs.\
(\ref{Veh}) and (\ref{Vee}), $V_{{\bf q}_1}(z)=\pm \frac{2\pi
e^2}{A \epsilon_0 q_1} {\rm e}^{-q_1 z}$ is the 2-dimensional
Fourier transform of the Coulomb interaction, where $A$ is the QW
area, and the plus (minus) sign is used in $V_{ee}$ ($V_{eh}$).
Using the anticommutation relations for the fermion operators, one
can easily find the scattering matrix elements to be
\begin{subequations}
\label{matel}
\begin{eqnarray}
\langle {\bf k}_{\rm x}+{\bf q} ; {\bf k}_{\rm e} - {\bf
q}|V_{eh}| {\bf k}_{\rm x} ; {\bf k}_{\rm e} \rangle &=&
\sum_{{\bf k}_1} \int dz_e dz_h dz_c \left[ V_{{\bf q}} (z_h -
z_c) \phi_{\alpha({\bf k}_{\rm x}+{\bf q})+ {\bf k}_1}(z_e,z_h)
\phi^*_{\alpha {\bf k}_{\rm x}+{\bf k}_1}(z_e,z_h) \psi_{{\bf
k}_{\rm e}-{\bf q}}
(z_c) \psi^*_{{\bf k}_{\rm e}}(z_c)- \right. \nonumber \\
&& \left. V_{{\bf k}_1+{\bf k}_{\rm e}-{\bf q}}(z_h - z_c)
\phi_{\alpha{\bf k}_{\rm x}+{\bf k}_1-\beta {\bf q}}(z_c,z_h)
\phi^*_{\alpha {\bf k}_{\rm x}+{\bf q}- {\bf k}_{\rm e}}(z_e,z_h)
\psi_{{\bf k}_{\rm e}-{\bf q}}(z_e) \psi^*_{{\bf k}_{\rm e}}(z_c)
\right]
\end{eqnarray}
\begin{eqnarray}
\langle {\bf k}_{\rm x}+{\bf q} ; {\bf k}_{\rm e} -{\bf q}
|V_{ee}| {\bf k}_{\rm x} ; {\bf k}_{\rm e}\rangle &=& \sum_{{\bf
k}_1} \int dz_e dz_h dz_c \left[ V_{{\bf q}}(z_e-z_c) \phi_{\alpha
{\bf k}_{\rm x}-\beta {\bf q}+ {\bf k}_1} (z_e,z_h) \phi^*_{\alpha
{\bf k}_{\rm x}+{\bf k}_1} (z_e,z_h) \psi_ {{\bf k}_{\rm e}-{\bf
q}}(z_c) \psi^*_{{\bf k}_{\rm e}}
(z_c) - \right. \nonumber \\
&& \left. V_{{\bf k}_1+{\bf k}_{\rm e}-{\bf q}} (z_e -z_c)
\phi_{\alpha {\bf k}_{\rm x}-\beta {\bf q}+ {\bf k}_1} (z_c,z_h)
\phi^*_{\alpha {\bf k}_{\rm x}+{\bf k}_1} (z_e,z_h) \psi_{{\bf
k}_{\rm e}-{\bf q}}(z_e) \psi^*_{{\bf k}_{\rm e}} (z_c) \right],
\end{eqnarray}
\end{subequations}
\end{widetext}
where we denote $\beta=1-\alpha$. The first term in each of
Eqs.~(\ref{matel}) contributes to the direct (classical) Coulomb
interaction and the second contributes to the exchange matrix
element. The direct term is calculated analytically (the details
can be found in Appendix A), resulting in:
\begin{equation}
V_{{\rm dir}}(q) = \frac{4 \pi e^2 \lambda^3}{\epsilon_0 A L^2}
h(\lambda q) \left[ g(\lambda \beta q/2) -g(\lambda \alpha q/2)
\right],
\label{Vdir}
\end{equation}
where $L$ is the QW width, $\lambda$ is a variational parameter
associated with the exciton Bohr radius in the QW, and we have
defined the dimensionless functions
\begin{eqnarray}
g(\lambda q)&=&\left[ 1+(\lambda q)^2 \right]^{-3/2} \nonumber \\
h(\lambda q)&=&\frac{{\rm e}^{-Lq}- 1+Lq + \frac{5(Lq)^3}{8\pi^2}
+\frac{3(Lq)^5}{32 \pi^4}}{(\lambda q)^3 \left[ 1+(Lq/2 \pi)^2
\right]^2}.
\label{gh}
\end{eqnarray}
It is evident from Eq.~(\ref{Vdir}) that the direct term is
identically zero for equal electron and hole masses
($\alpha=\beta=0.5$). Similarly, we have for the exchange term
\begin{eqnarray}
&V\!\!\!\!&_{\rm exc}(\Delta {\bf k},{\bf q}) =-\frac{8e^2
\lambda^5}{ \epsilon_0 AL^2} \int d^2 k_1 g(\lambda |{\bf k}_1+
\alpha {\bf q}-\Delta {\bf k}|) \times \nonumber
\\ && \left[ g(\lambda|{\bf k}_1+{\bf
q}-\Delta {\bf k}|)-g(\lambda |{\bf q}- \Delta {\bf k}|) \right]
h(\lambda k_1) ,
\label{Vexch}
\end{eqnarray}
where we have defined $\Delta {\bf k}={\bf k}_{\rm e}-\alpha {\bf
k}_{\rm x}$. The exchange term (\ref{Vexch}) is computed
numerically, however its angular part can be calculated
analytically (see Appendix B). It is convenient to transform to
dimensionless direct and exchange integrals given by
\[
V=\frac{2}{\pi} \frac{e^2 \lambda^3}{\epsilon_0 AL^2}I.
\]
The direct and exchange integrals are plotted in
Fig.~\ref{dir_exc} as a function of the transferred momentum $q$,
for the case $\Delta {\bf k}=0$, where the angular dependence of
$I_{\rm exc}$ disappears. As $q \rightarrow 0$ the direct integral
approaches zero , while the exchange integral has its maximum
(this is also the case for exciton-exciton interaction
\cite{Ciuti1}). In the general case, the exchange integral is a
function of the transferred momentum $q$, the momentum difference
$\Delta k$ (which can be regarded, for convenience, as the
in-plane momentum of the colliding electron in the rest frame of
the exciton), and of the angle $\theta = \angle ({\bf q},\Delta
{\bf k})$. The exchange interaction term has the following
features: (i) The interaction favors the case ${\bf q}=\Delta {\bf
k}$. Physically, this means that the electron is inclined to
transfer as much momentum as possible to the exciton, preferably
in the same direction. (ii) The interaction retains its strength
for quite large values of $q$ (or $\Delta k$) even though the
excitonic wave function vanishes much more rapidly with momentum.
(iii) The differential cross section is largest for $\theta=0$ and
decreases to a minimum for back scattering ($\theta=\pi$).

\vspace*{2.2cm}

Next we calculate the scattering rate of an exciton, with an
initial in-plane momentum ${\bf k}_{\rm x}$, to an excitonic state
with an in-plane momentum ${\bf k}_{\rm x}+{\bf q}$. The exchange
matrix element (\ref{Vexch}) depends on five variables: $k_{\rm
x},k_{\rm e},q, \cos \theta,\cos \phi$, where we have defined
$\phi =\angle({\bf k}_{\rm x},{\bf k}_{\rm x}+{\bf q})$. It is
convenient to transform from the angles $\theta$ and $\phi$ to the
angles $\gamma=\angle({\bf k}_{\rm x},{\bf q})$ and
$\delta=\angle({\bf k}_{\rm e},{\bf q})$ \cite{delta}:
\begin{eqnarray}
\cos \theta &=& \frac{k_{\rm e} \cos \delta -\alpha k_{\rm x} \cos
\gamma}{\sqrt{k_{\rm e}^2+\alpha^2 k_{\rm x}^2-2 \alpha k_{\rm x}
k_{\rm e} \cos(\gamma \pm \delta)}}, \ \ (\Delta k \neq 0) \nonumber \\
\cos \phi &=& \frac{k_{\rm x}+q \cos \gamma}{\sqrt{k_{\rm
x}^2+q^2+2k_{\rm x}q \cos \gamma}}, \ \ ({\bf k}_{\rm x} \neq
-{\bf q}).
\label{gamdel}
\end{eqnarray}
It can be verified that for the special case of $k_{\rm x}=0$, we
have $\theta=\delta$ and $\phi=\gamma$. Conservation of energy and
momentum for the exciton-electron scattering process reads
\begin{equation}
E_{\rm x}(\left|{\bf k}_{\rm x}+{\bf q} \right|)+E_{\rm e}(|{\bf
k}_{\rm e}-{\bf q}|)=E_{\rm x}(k_{\rm x})+E_{\rm e}(k_{\rm e}),
\label{Econs}
\end{equation}
where $E_{\rm x}$ ($E_{\rm e}$) is the exciton (electron) kinetic
energy. This equation is satisfied with:
\begin{equation}
k_{\rm e}^* =\frac{1}{2M_{\rm x} \cos \delta} \left[ q(M_{\rm
x}+m_{\rm e})+2k_{\rm x} m_{\rm e} \cos \gamma \right] \ ,
\end{equation}
The scattering rate is then found using Fermi's golden rule
\begin{widetext}
The scattering rate is then found using Fermi's golden rule
\begin{eqnarray}
w_{\rm x-e}({\bf k}_{\rm x} \rightarrow {\bf k}_{\rm x}+{\bf q})
&=& \frac{A}{\pi \hbar} \int d^2 k_{\rm e} \frac{| V_{\rm
dir}+V_{\rm exc}|^2 \delta (k_{\rm e}-k_{\rm e}^*)}{\left|
\frac{dE_{\rm e}(|{\bf k}_{\rm e}-{\bf q}|)}{dk_{\rm e}} -\frac{d
E_{\rm e}(k_{\rm e})}{dk_{\rm e}} \right|}_{k_{\rm e}=k_{\rm e}^*}
f_{\rm fd} (k_{\rm e})\left[1-f_{\rm fd} (|{\bf k}_{\rm e}-{\bf
q}| ) \right] \nonumber \\
&=& \frac{A m_e}{\pi \hbar^3} \int_0^{2 \pi} \frac{d \delta}
{\left| \cos \delta \right|}\frac{k_{\rm e}^*}{q} \left|V_{\rm
dir} + V_{\rm exc} \right|^2
 f_{\rm fd} (k_{\rm e}^*)\left[1-f_{\rm fd} (|{\bf k}_{\rm
e}^*-{\bf q}| ) \right],
\label{rate_ex}
\end{eqnarray}
\end{widetext}
where $f_{\rm fd}(k_{\rm e})=\left[ {\rm e}^{(E_{\rm e}(k_{\rm
e})-\mu)/k_B T}+1 \right]^{-1}$ is the electron Fermi-Dirac
distribution function. In the two dimensional case that we are
considering, the chemical potential is $\mu=k_B T \log ({\rm
e}^{E_f/k_B T}-1)$, and we assume that the Fermi energy is given
by the free 2D electrons value: $E_f=\pi \hbar^2 n_{\rm e}/m_e$.
We note that in the high temperature limit, $E_f \ll k_B T$, the
Fermi-Dirac distribution is practically classical, and that for
low electron densities, the chemical potential becomes negative.

Using the relations (\ref{gamdel}), the integrand in
(\ref{rate_ex}) can be expressed as a function of the variables:
$k_{\rm x},q,\cos \gamma, \cos \delta$. The exciton linewidth due
to electron scattering is calculated by integrating over all final
excitonic states
\begin{equation}
\Gamma_{\rm x-e} (k_{\rm x}) = \frac{\hbar A}{(2 \pi)^2} \int q dq
d \gamma w_{\rm x-e}(k_{\rm x},q,\cos \gamma) \ . \label{Gam_ex}
\end{equation}

\begin{figure}[htbp]
\epsfxsize=0.55\textwidth
\centerline{\epsffile{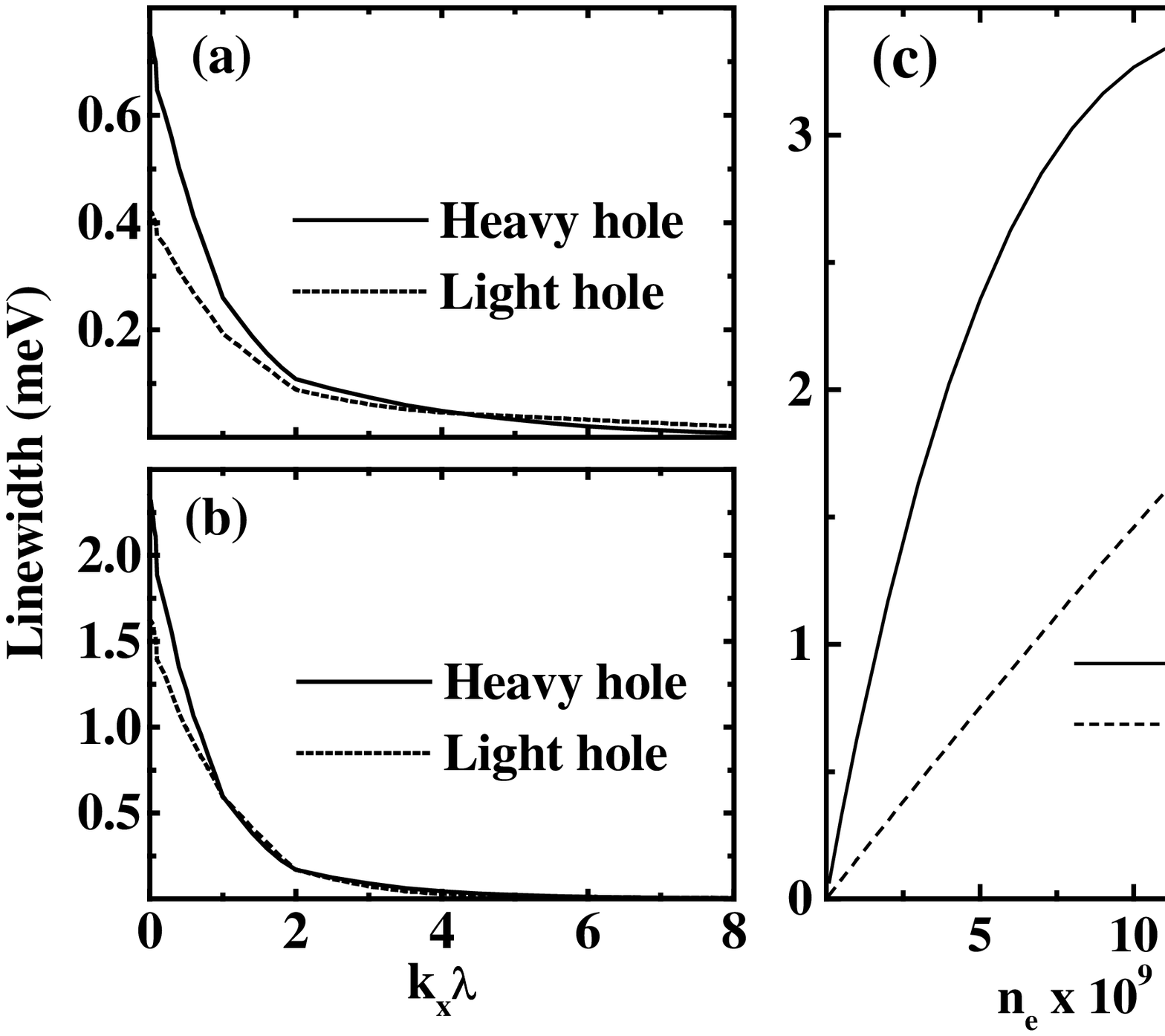}}
\vspace*{-4cm}
\caption{(a) Heavy exciton ($M_{\rm x}=0.177m_e$) and light
exciton ($M_{\rm x}=0.306m_e$) linewidths due to electron
scattering for a 2DEG with $n_{\rm e}=5 \cdot 10^9 {\rm cm}^{-2}$,
as a function of the exciton initial in-plane momentum, at $T=80K$
; (b) Same as (a) at $T=5K$ ; (c) Linewidth of heavy hole exciton
with initial momentum $k_{\rm x}=0$, as a function of $n_{\rm e}$,
for the two temperatures considered.}
\label{fgam_ex}
\end{figure}

Figures \ref{fgam_ex} a,b show the heavy-hole exciton ($M_{\rm
x}=0.177m_e$) and light-hole exciton ($M_{\rm x}=0.306m_e$)
linewidths at $T=80K$ and $T=5K$ for $n_{\rm e}=5\cdot 10^9 {\rm
cm}^{-2}$. The large linewidths obtained for relatively low
$n_{\rm e}$ reflect the high efficiency of the electron scattering
mechanism. This should be compared to an exciton linewidth of
$\approx$ 0.2meV for acoustic phonon scattering at $T=80K$. The
larger linewidth obtained at $T=5K$ (Fig. \ref{fgam_ex}b) is
explained by noticing that the exciton-electron interaction matrix
elements favor small energy transfer transitions. Thus, at high
temperatures, the electrons have too high energy to be effective
scatterers. Decreasing $T$ thus increases the scattering rate
until the electron gas becomes degenerate, and fewer final states
are available for the scattered electrons. In figure
\ref{fgam_ex}c the heavy hole exciton linewidth at $k_{\rm x}=0$
is plotted as a function of $n_{\rm e}$ for $T=5K$ and $80K$ (the
light hole exciton linewidth exhibits a similar behaviour). It is
seen that while at $T=80K$ the linewidth amplitudes scale
practically linearly with $n_e$ for a large range of electron
densities (up to $n_{\rm e} \approx 10^{11} {\rm cm}^{-2}$), the
linearity region at $T=5K$ is much smaller. This linearity
threshold seems to be in accord with the onset of the phase-space
filling effect which becomes noticeable at higher densities and
effectively enlarges the exciton Bohr radius \cite{Huang}. As the
temperature decreases, the effect of phase-space filling becomes
important at a much lower electron density. We note that the
functional dependence of the linewidths on the initial exciton
momentum is almost unchanged within the linear regime. Increasing
$n_{\rm e}$ further results in a shift of the maximum linewidth
from $k_{\rm x}=0$ to higher momenta.
\vspace*{2cm}

\subsection{Exciton Dissociation scattering}

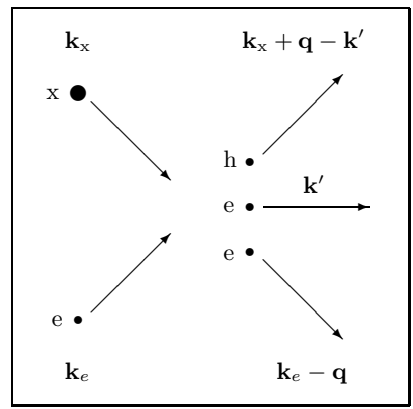
\begin{figure}[htbp]
\begin{picture}(150,150)
\put(0,0){\framebox(150,150)}

\put(13,116){x} \put(25,118){\circle*{6}}
\put(30,115){\vector(1,-1){30}}

\put(20,135){${\bf k}_{\rm x}$}

\put(15,30){e} \put(25,32){\circle*{3}}
\put(30,35){\vector(1,1){30}}

\put(20,10){${\bf k}_e$}

\put(80,90){h} \put(90,92){\circle*{3}}
\put(95,95){\vector(1,1){30}}

\put(87,135){${\bf k}_{\rm x}+{\bf q}-{\bf k}'$}

\put(80,73){e} \put(90,75){\circle*{3}}
\put(95,75){\vector(1,0){40}}

\put(110,80){${\bf k}'$}

\put(80,55){e} \put(90,58){\circle*{3}}
\put(95,55){\vector(1,-1){30}}

\put(100,10){${\bf k}_e-{\bf q}$}

\end{picture}
\caption{Electron-exciton inelastic scattering scheme}
\label{xion_scat}
\end{figure}

We now consider the case when the electron-exciton scattering
results in dissociation. This scattering process is depicted
schematically in Fig.\ \ref{xion_scat}.
The initial state of the system is given again by
Eq.~(\ref{initial}) while the final state is that of three free
particles, symbolically written  as $|\bf{ k}_{\rm e}-{\bf q} ;
{\bf k}' ; {\bf k}_{\rm x}+{\bf q}-{\bf k}' \rangle$, where ${\bf
q}$ is the momentum transferred from the free electron to the
electron-hole pair, and ${\bf k}'$ is the second electron
momentum. Calculating the Coulomb interaction matrix elements
between the initial and final states of the system, we again
obtain direct and exchange interaction terms:
\begin{subequations}
\begin{eqnarray}
V_{{\rm dir}}^{\rm dis}\!\!\!&(&\!\!\!{\bf k}_{\rm x},{\bf q},{\bf
k}') = \frac{4 e^2 \lambda^4}{\epsilon_0 L^2}\left(\frac{2
\pi}{A}\right)^{3/2} h(\lambda q) \times \nonumber \\
&& \left[ g(\lambda|\alpha {\bf k}_{\rm x}+{\bf q}-{\bf k}'|)-
g(\lambda|\alpha {\bf k}_{\rm x}-{\bf k}'|)
\right] \\
V_{{\rm exc}}^{\rm dis}\!\!\!&(&\!\!\!{\bf k}_{\rm x},{\bf q},{\bf
k}',{\bf k}_{\rm e}) = -\frac{4 e^2 \lambda^4}{\epsilon_0
L^2}\left(\frac{2 \pi}{A}\right)^{3/2} h(\lambda \left|{\bf
k}'-{\bf k}_{\rm e}\right|) \times \nonumber \\
&&  \left[ g(\lambda|\alpha {\bf k}_{\rm x}+{\bf q}-{\bf k}'|)
-g(\lambda|\alpha {\bf k}_{\rm x}+{\bf q}-{\bf k}_{\rm e}|)
\right]
\end{eqnarray}
\end{subequations}
where $g(\lambda q),h(\lambda q)$ are given in (\ref{gh}). We note
that contrary to the elastic scattering case, here the direct and
exchange matrix elements are comparable. In order to calculate the
scattering rate we first introduce energy-momentum conservation
\begin{eqnarray}
E_{\rm x}(k_{\rm x})+E_{\rm e}(k_{\rm e})&=&E_{\rm b}+E_{\rm
h}(|{\bf k}_{\rm x}+{\bf q}-{\bf k}'|)+ \nonumber \\
&& E_{\rm e}(k')+E_{\rm e}(|{\bf k}_{\rm e}-{\bf q}|),
\end{eqnarray}
where $E_{\rm b}$ is the exciton binding energy. Denoting the
angle $\nu=\angle({\bf k}_{\rm x},{\bf k}')$ together with
$\gamma,\delta$ which were defined previously, this equation is
satisfied with
\begin{eqnarray}
&k\!\!&\!\!_{\rm e}^*=\frac{1}{2\beta q\cos \delta} \left\{
\frac{2
\beta E_{\rm b}m_{\rm e}}{\hbar^{2}}+(\alpha k_{\rm x})^{2}+q^{2}+k^{\prime 2}+
\right. \nonumber \\
&&\left. 2\alpha \left[k_{\rm x}q \cos \gamma-k'k_{\rm x}\cos
\nu-k'q \cos(\gamma-\nu) \right] \right\}
\end{eqnarray}
The scattering rate is then calculated to be:
\begin{eqnarray}
&\!\!\!w\!\!\!&_{\rm x-e}^{\rm dis}({\bf k}_{\rm x}
\rightarrow{\bf k}_{\rm x}+{\bf q}) = \frac{A^2m_{\rm e}}{4
\pi^3\hbar^3} \int \frac{k' dk' d\nu d\delta}{|\cos \delta|}
\frac{k_{\rm e}^*}{q} \times \nonumber \\ && \left| V_{\rm
dir}^{\rm dis} \pm V_{\rm exc}^{\rm dis} \right|^2
 f_{\rm fd}(k_{\rm e}^*) \left[ 1-f_{\rm fd}(|{\bf
k}_{\rm e}^*-{\bf q}|) \right] \times \\
&&\left[ 1-f_{\rm fd}(k') \right] \nonumber,
\label{xrate_ion}
\end{eqnarray}
where the + (-) sign between the direct and exchange matrix
elements corresponds to the singlet (triplet) electron spin
configuration. For an unpolarized electron gas, an averaging over
the two spin configurations must be performed, in order to take
into account the contributions of all the electrons to the exciton
linewidth due to dissociating scattering:
\begin{equation}
\Gamma_{\rm x-e}^{\rm dis}(k_{\rm x})=\frac{3}{4}\Gamma_{\rm
x-e}^{\rm dis,-}(k_{\rm x})+\frac{1}{4} \Gamma_{\rm x-e}^{\rm
dis,+}(k_{\rm x}).
\end{equation}
$\Gamma_{\rm x-e}^{\rm dis,+}(k_{\rm x})$ ($\Gamma_{\rm x-e}^{\rm
dis,-}(k_{\rm x})$) denotes the exciton linewidth contribution
from the singlet (triplet) spin configuration, and both are
calculated using Eq.~(\ref{Gam_ex}). We note that the spin
configuration averaging is irrelevant in the elastic scattering
case, since the direct term there is much smaller than the
exchange term. This results in similar contributions from both
singlet and triplet configurations.

\vspace*{3.3cm}
\begin{figure}[htbp]
\epsfxsize=0.65\textwidth
\centerline{\epsffile{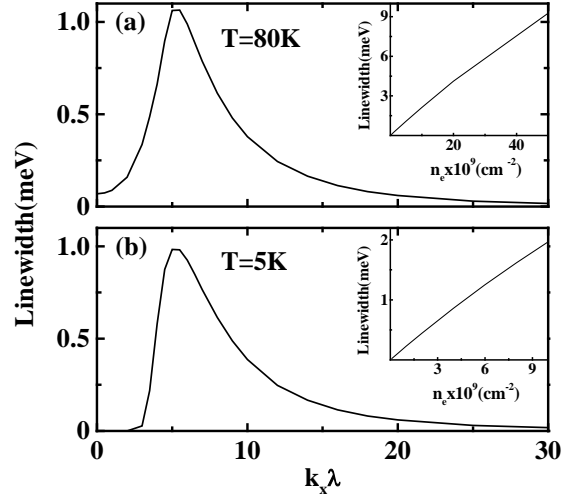}}
\vspace*{-4.6cm}
\caption{Heavy exciton ($M_{\rm x}=0.177m_e$) linewidths
due to dissociating scattering with electrons, as a function of
exciton initial in-plane momentum, with $n_{\rm e}=5\cdot 10^{9}
{\rm cm}^{-2}$, at (a) $T=80K$ ; (b) $T=5K$. The insets show the
maximum linewidth dependence on $n_{\rm e}$.}
\label{lw_ion}
\end{figure}

Figure \ref{lw_ion}a shows the heavy hole exciton linewidth due to
dissociating scattering at $n_{\rm e}=5\cdot 10^{9} {\rm
cm}^{-2}$, and $T=80K$. Although the magnitude of $\Gamma_{\rm
x-e}^{\rm dis} (k_{\rm x})$ is of the same order as the elastic
scattering linewidth (compare to Fig. \ref{fgam_ex}a), its
functional dependence on the exciton in-plane momentum is very
different. In particular, we note that the maximal linewidth is
obtained at a very large momentum ($k_{\rm x} \approx 5
\lambda^{-1}$). This is due to the fact that in order for an
exciton with initially small $k_{\rm x}$ to be ionized, it must
scatter on an electron with energy large enough to overcome its
binding energy. This is less likely as the temperature decreases,
as is evident from Fig.\ \ref{lw_ion}b, where $\Gamma_{\rm
x-e}^{\rm dis} (k_{\rm x})$ is plotted at $T=5K$.  The insets in
Figure \ref{lw_ion} show the maximum values of $\Gamma_{\rm
x-e}^{\rm dis} (k_{\rm x})$ vs. $n_{\rm e}$. As in the elastic
scattering case, the linear dependence on $n_{\rm e}$ holds for
much larger electron densities at $T=80K$.

\section{Charged exciton (trion)-electron scattering}
\label{trion}
\setcounter{equation}{0}

At low temperatures and low $n_{\rm e}$, charged excitons are
formed. In this section we calculate the trion linewidth due to
scattering with free electrons. Three scattering processes are
considered: elastic scattering, where the trion remains bound,
capturing scattering, where the heavy-hole exciton captures a free
electron to form a trion, and dissociating scattering where the
trion dissociates into a heavy-hole exciton and an extra free
electron. The latter process is important due to the small binding
energy of the trion with respect to the exciton. We neglect the
scattering in which both electrons become unbound. In order to
facilitate the matrix elements calculation, we consider a simple
two parameter Chandrasekhar type \cite{Chan} trial function for
the trion
\begin{eqnarray}
&\phi\!\!&\!\!^{\rm tr}(r_{1{\rm h}},r_{2{\rm
h}},z_1,z_2,z_h)=\frac{1}{A} {\cal N}_{\rm tr} \chi_e(z_1)
\chi_e(z_2) \times \nonumber \\
&&\chi_h(z_h) \left[ \phi (r_{1{\rm h}}) \phi'(r_{2{\rm h}}) \pm
\phi' (r_{1{\rm h}}) \phi(r_{2{\rm h}}) \right] ,
\label{phitr}
\end{eqnarray}
where the + (-) sign applies to the singlet (triplet) spin
configuration. In Eq.\ (\ref{phitr}), $r_{\rm ih}=|{\bf r}_{\rm
i}-{\bf r}_{\rm h}|, \ (i=1,2)$ are the in-plane coordinates of
the two electrons with respect to the hole in-plane coordinate,
$z_i$ are the electrons $z$ coordinates, and $\chi_e(z_i)$,
$\chi_h(z_h)$ are the confinement functions in the $z$ direction,
taken to be the same for all constituents (Eq.\ (\ref{chi})). The
two electron orbitals are given by the normalized functions
\begin{subequations}
\begin{eqnarray}
\phi(r)&=&\sqrt{\frac{2}{\pi \lambda^2}} {\rm e}^{-r/\lambda} \\
\phi'(r)&=&\sqrt{\frac{2}{\pi \lambda^{\prime 2}}} {\rm
e}^{-r/\lambda'},
\end{eqnarray}
\end{subequations}
and the trion wavefunction normalization factor is given by
\begin{equation}
{\cal N}_{\rm tr}=\left( \frac{2}{L}\right)^{3/2} \frac{1}{\sqrt{1
\pm \kappa^2}},
\end{equation}
where
\begin{equation}
\kappa \equiv \langle \phi|\phi' \rangle=\frac{4 \lambda
\lambda'}{(\lambda+\lambda')^2}.
\label{s}
\end{equation}
The trial function (\ref{phitr}) was used in \cite{SanPro} for the
case of a two dimensional negative-donor center (i.e., $m_h
\rightarrow \infty$ limit). In addition, these authors performed
the binding energy calculations using a more complicated version
of (\ref{phitr}) which includes a correlation term (with an
additional variational parameter). The variational parameters
$\lambda,\lambda'$ are calculated by maximizing the trion binding
energy for the two spin configurations. The details of the binding
energy calculation are given in Appendix C, resulting in $E_b^{\rm
tr}=0.985 {\rm meV}, \lambda=143$\AA, $\lambda'=300$\AA ($E_b^{\rm
tr} =0.765 \mu {\rm eV}, \lambda=151$\AA, $\lambda'=7650$\AA) for
the singlet (triplet) spin configuration, where $E_b^{\rm tr}$ is
calculated with respect to the heavy hole 1S neutral exciton
energy. We note that the triplet configuration is barely bound, in
accordance with the experimental observations, thus it can be
disregarded. Comparing with the much more elaborate treatment of
St\`{e}b\`{e} {\it et al} \cite{Stebe}, our results are fairly
accurate, giving us the confidence to proceed in calculating the
trion-electron scattering matrix elements using the wave function
(\ref{phitr}).

We construct the state of a single trion with an in-plane CM
momentum ${\bf k}_t$, similarly to that of a single exciton
(\ref{ex_st}):
\begin{eqnarray}
| {\bf k}_{\rm t} \rangle &=& \sum_{\stackrel{{\bf k}_1,{\bf
k}_2}{s_1,s_2}} \int dz_1 dz_2 dz_h \phi^{\rm tr*}_{\alpha_t {\bf
k}_{\rm t}+{\bf k}_1,\alpha_t {\bf k}_{\rm t}+{\bf k}_2}
(z_1,z_2,z_h) \times \nonumber \\
&& \xi^*_S(s_1,s_2) c^{s_1\dag}_{-{\bf k}_1,z_1}
c^{s_2\dag}_{-{\bf k}_2,z_2}d^\dag_{{\bf k}_{\rm t}+{\bf k}_1+{\bf
k}_2,z_h} | 0 \rangle ,
\label{tr_st}
\end{eqnarray}
where we have added the spin index to the electron creation
operators, and denoted $\alpha_t=m_{\rm e}/M_{\rm tr}$, and
$\xi_S(s_1,s_2)=\langle S|s_1,s_2\rangle$ as the projection of a
generic spin configuration of two electrons on the singlet spin
configuration. The in-plane Fourier transformed trion wave
function in Eq.\ (\ref{tr_st}) is given by
\begin{equation}
\phi^{\rm tr}_{{\bf k}_1,{\bf k}_2}(z_1,z_2,z_h)={\cal N}_{\rm tr}
\chi_e(z_1) \chi_e(z_2) \chi_h(z_h) \phi_{{\bf k}_1}\phi'_{{\bf
k}_2},
\label{tr_k}
\end{equation}
where,
\begin{subequations}
\begin{eqnarray}
\phi_{{\bf k}}&=&\sqrt{\frac{8 \pi \lambda^2}{A}} \frac{1}{
\left[1+(\lambda k)^2 \right]^{3/2}} \\
\phi'_{{\bf k}}&=&\sqrt{\frac{8 \pi \lambda^{\prime 2}}{A}}
\frac{1}{ \left[1+(\lambda' k)^2 \right]^{3/2}}.
\end{eqnarray}
\end{subequations}
A state comprising a trion and a free electron having ${\bf
k}_{\rm e}$, $z_e$, and $s_e$ will be written as
\begin{eqnarray}
| {\bf k}_{\rm t} \!\!\!&\!\!;\!\!\!& {\bf k}_{\rm e}
\rangle=\sum_{\stackrel{{\scriptstyle {\bf k}_1,{\bf
k}_2}}{s_1,s_2,s_e}} \int dz_1 dz_2 dz_h dz_e \xi^*_S(s_1,s_2)
\times \nonumber \\
&& \phi^{\rm tr*}_{\alpha_t {\bf k}_{\rm t}+{\bf k}_1,\alpha_t
{\bf k}_{\rm t}+{\bf k}_2} (z_1,z_2,z_h) \psi^*_{{\bf k}_{\rm e}}
(z_e) \times \label{initr} \\
&& c^{s_1\dag}_{-{\bf k}_1,z_1} c^{s_2\dag}_{-{\bf
k}_2,z_2}d^\dag_{{\bf k}_{\rm t}+{\bf k}_1+{\bf k}_2,z_h}
c^{s_e\dag}_{{\bf k}_{\rm e},z_e}| 0 \rangle. \nonumber
\end{eqnarray}
In the following we shall use these wave functions to calculate
the scattering matrix elements for the various scattering
processes.

\subsection{Elastic scattering}

Applying the Coulomb interaction operators $V_{e1}$, $V_{e2}$, and
$V_{eh}$ to the initial state of the system (\ref{initr}), and
discarding the self energy contributions, we can write a generic
expression for the three scattering matrix elements:
\begin{widetext}
\begin{eqnarray}
\langle {\bf k}_{\rm t}\!\!&+&\!\!{\bf q} ; {\bf k}_{\rm e}-{\bf
q}|V_{ei}| {\bf k}_{\rm t} ; {\bf k}_{\rm e} \rangle = \frac{1}{2}
\sum_{\stackrel{\scriptstyle{{\bf k}_1,{\bf k}_2}}{{\bf k}'_1,{\bf
k}'_2,{\bf q}_1}}
\sum_{\stackrel{\scriptstyle{s_1,s_2,s_e}}{s'_1,s'_2,s'_e}} \int
dz_i dz'_i dz_h dz'_h V_{{\bf q}_1}(z_e-z_i) \phi^{\rm
tr}_{\alpha_t ({\bf k}_{\rm t}+{\bf q})+{\bf k}'_1, \alpha_t ({\bf
k}_{\rm t}+{\bf q})+{\bf k}'_2} (z'_i) \times \nonumber \\
&& \xi_S(s'_1,s'_2) \psi_{{\bf k}_{\rm e}-{\bf q}} (z'_e)
\phi^{\rm tr*}_{\alpha_t {\bf k}_{\rm t}+{\bf k}_1,\alpha_t {\bf
k}_{\rm t}+{\bf k}_2} (z_i) \xi^*_S(s_1,s_2) \psi^*_{{\bf k}_{\rm
e}} (z_e) \langle 0|c_1 c_2 c_3 c_4^\dag c_5^\dag c_6^\dag |0
\rangle, \ \ (i=1,2,h),
\label{vei}
\end{eqnarray}
\end{widetext}
where we put the $1/2$ prefactor to indicate averaging over the
initial electron spin states, and used a numeric index for the
electron operators, indicating both spatial and spin degrees of
freedom. Note that we have omitted the hole operators since they
anticommute with the electron operators and do not contribute
additional constraints. Using the electrons anticommutation
relations, the Fermi vacuum expectation value of the operators
reads:
\begin{eqnarray}
\langle 0|c_1 c_2 c_3 c_4^\dag c_5^\dag c_6^\dag |0 \rangle &=&
\delta_{34}(\delta_{25}\delta_{16}-\delta_{15}\delta_{26})+\nonumber
\\ \delta_{24}
(\delta_{15}\delta_{36}-\delta_{35}\delta_{16})&+&\delta_{14}
(\delta_{35}\delta_{26}-\delta_{25}\delta_{36}),
\label{delspin}
\end{eqnarray}
where we have symbolically written $\delta_{ij}=\delta_{{\bf
k}_i,{\bf k}_j}\delta(z_i-z_j)\delta_{s_i,s_j}$. For the 1st
electron interaction term we have:
\begin{equation}
\begin{array}{ll}
1=({\bf k}_e-{\bf q} ; z'_e ; s'_e) \ \ & 2=(-{\bf k}'_2
; z'_2 ; s'_2) \\
3=(-{\bf k}'_1 ; z'_1 ; s'_1) \ \ & 4=(-{\bf k}_1-{\bf q}_1 ; z_1
; s_1) \\ 5=(-{\bf k}_2 ; z_2 ; s_2) \ \ & 6=(-{\bf k}_e+{\bf q}_1
; z_e ; s_e)
\end{array}
\label{ops}
\end{equation}
The second electron interaction term is found by making the
substitutions: ${\bf k}_1 \rightarrow {\bf k}_1-{\bf q}_1$, ${\bf
k}_2 \rightarrow {\bf k}_2+{\bf q}_1$ in Eqs.\ (\ref{ops}).
Similarly, the electron-hole interaction term is found by
substituting: ${\bf k}_1 \rightarrow {\bf k}_1-{\bf q}_1$ in
(\ref{ops}). Applying the delta functions of (\ref{delspin}) to
Eq.\ (\ref{vei}), it is evident that the contributions of the
$z$-dependent parts are identical for all three terms, since we
have taken identical confinement functions for the electron and
the hole (see Eq.\ \ref{chi}). Assuming that the trion remains in
the singlet electron spin configuration, we sum over the spin
degrees of freedom, resulting in selection rules for the three
electron final spin states, which determine the relevant signs of
the various terms. Performing the in-plane momentum integrations,
and collecting the matrix elements into one direct and two
exchange contributions, we finally find:
\begin{widetext}
\begin{eqnarray}
V_{\rm dir}(q)&=& {\cal N}_{\rm es} h(\lambda q)\left\{ \left[
g(\lambda \alpha_t q/2) g(\lambda' \beta_t q/2)+g(\lambda'
\alpha_t q/2) g(\lambda \beta_t q/2)+ 2\kappa^2g(\tilde{\lambda}
\alpha_t q) g(\tilde{\lambda}\beta_t q) \right]-
\right. \nonumber \\
&& \left. \left[ g(\lambda \alpha_t q/2) g(\lambda' \alpha_t
q/2)+\kappa^2g^2(\tilde{\lambda} \alpha_t q) \right] \right\}
\label{trion_dir}
\end{eqnarray}
and
\begin{subequations}
\label{trion_exc}
\begin{eqnarray}
V^{(1)}_{\rm exc}({\bf q},\Delta{\bf k}_t)&=&-\frac{{\cal N}_{\rm
es}}{\pi} \int d^2 k_1 h(\lambda k_1)\left\{ \left[ \lambda
\lambda' \kappa g(\tilde{\lambda} \alpha_t q) g(\lambda |{\bf
k}_1-\Delta {\bf k}_t+\alpha_t {\bf q}|) + \lambda^{\prime 2}
g(\lambda \alpha_t q/2)g(\lambda' |{\bf k}_1-\Delta {\bf
k}_t+\alpha_t {\bf
q}|) \right] \times \right. \nonumber \\
&& \left.  \left[ g(\lambda' |{\bf k}_1-\Delta {\bf k}_t+{\bf
q}|)- g(\lambda' |\Delta {\bf k}_t- {\bf q}|) \right] + [\lambda
\leftrightarrow \lambda']\times [\lambda \leftrightarrow
\lambda'] \right\} \\
V^{(2)}_{\rm exc}({\bf q},\Delta{\bf k}_t)&=& -\frac{{\cal N}_{\rm
es}}{\pi} \int d^2 k_1 h(\lambda k_1)\left\{ \left[ \lambda
\lambda' \kappa g(\tilde{\lambda}|{\bf k}_1-\alpha_t q|)
g(\lambda' |{\bf q}-\Delta {\bf k}_t|) + \lambda^2 g(\lambda'
|{\bf k}_1-\alpha_t {\bf q}|/2) g(\lambda |{\bf q}-\Delta {\bf
k}_t|) \right] \times \right. \nonumber \\
&& \left. g(\lambda |{\bf k}_1-\Delta {\bf k}_t+\alpha_t{\bf q}|)
+ [\lambda \leftrightarrow \lambda']\times[\lambda \leftrightarrow
\lambda'] \right\},
\end{eqnarray}
\end{subequations}
\end{widetext}
where we have denoted $\beta_t=1-\alpha_t$, $\Delta {\bf k}_t={\bf
k}_e-\alpha_t {\bf k}_t$, and $\tilde{\lambda}=\lambda
\lambda'/(\lambda+\lambda')^2$. In Eqs.\
(\ref{trion_dir})-(\ref{trion_exc}), the prefactor, multiplying
the dimensionless quantities, is given by
\begin{equation}
{\cal N}_{\rm es}=\frac{1}{1+\kappa^2}\frac{4 \pi e^2
\lambda^3}{L^2 A\epsilon_0}.
\end{equation}
We note that the direct term (\ref{trion_dir}) diverges at $q
\rightarrow 0$. This divergence originates from the infinite range
of the Coulomb potential, which is manifest whenever a scattering
event between two charged particles occurs (note that for the
neutral exciton this divergence cancelled out, due to the equal
contributions with opposite signs from the electron and hole
constituting the exciton). In practice the Coulomb potential is
screened by the presence of the 2DEG. Since we are considering a
low density 2DEG, its Fermi wavevector, $k_{\rm f}=\sqrt{2\pi
n_{\rm e}}$ is small, thus the semiclassical (Thomas-Fermi)
approximation, which assumes that $q \ll k_{\rm f}$, is inadequate
\cite{Bastard}. In particular, the two-dimensional semiclassical
analysis results (at $T=0$) in a density-independent screening
wavevector $q_0$, implying that the presence of a single electron
is sufficient to screen the external potential, which is clearly
unrealistic. We adopt instead the Lindhard approach which utilizes
a perturbative scheme to evaluate the induced charge density in
first order of the total potential (By total we mean the sum of
the external potential, i.e., that of the trion, and the 2DEG
potential that is induced by the trion's presence). The in-plane
Fourier transformed dielectric function which results from the
effect of the screening is given by \cite{Bastard}
\begin{equation}
\epsilon_s(q)=1+\frac{q_0}{q} g_s(q) \alpha_s(q)
\label{screene}
\end{equation}
where
\begin{equation}
q_0=\frac{2 m_e e^2}{\epsilon_0 \hbar^2}=\frac{2}{a_{\rm B}}
\end{equation}
and $a_{\rm B}$ is the electrons effective bulk Bohr radius. Two
factors appear in Eq.\ (\ref{screene}), which influence the
screening. The first is the screening form factor given by
\begin{equation}
g_s(q)=\int dz dz' \chi^2(z)\chi^2(z'){\rm e}^{-q |z-z'|}
\label{sff}
\end{equation}
which arises from the finite QW width. In the infinite barrier
limit we are considering, the confinement functions, $\chi(z)$,
are given by Eq. (\ref{chi}), and $g_s(q)$ can be calculated
analytically, resulting in
\begin{equation}
g_s(q)=\frac{2}{(qL)^2}\frac{{\rm e}^{-Lq}-1+Lq+\frac{5}{8\pi^2}
(Lq)^3+\frac{3}{32\pi^4}(Lq)^5}{\left[1+\left(\frac{Lq}{2\pi}\right)^2
\right]^2}.
\end{equation}
This screening form factor is smaller than 1, thus it reduces the
effect of the screening, compared with the strictly
two-dimensional case ($g_s(L\rightarrow 0)=1$). The second factor
which appears in Eq. (\ref{screene}) is given by
\begin{equation}
\alpha_s(q)=\frac{1}{\pi} \int d^2k \frac{f_{\rm fd}({\bf k}+{\bf
q})-f_{\rm fd}(k)}{({\bf k}+{\bf q})^2-k^2}.
\end{equation}
We note that $\alpha_s(q)$ does not depend on the particular
details of the QW, but only on the two-dimensional nature of the
2DEG motion. Substituting the screening dielectric function into
the Coulomb interaction, we find the screened potential to be
\begin{equation}
V_q^s (z)=\frac{1}{\epsilon_s (q)} \frac{2 \pi e^2}{A \epsilon_0}
\frac{e^{-|z|q}}{q} = \frac{2 \pi e^2}{A \epsilon_0}
\frac{e^{-|z|q}}{q+q_0 g_s(q) \alpha_s(q)} ;
\end{equation}
thus the use of the screened potential implies replacing
$h(\lambda q)$ in Eq.\ (\ref{trion_dir}) with
\begin{equation}
h_s(\lambda q)=\frac{h(\lambda q)}{1+\frac{4 \lambda^3}{L^2 a_{\rm
B}}h(\lambda q) \alpha_s(q)} .
\end{equation}
In general, the effect of the quasi two-dimensional screening
saturates in larger electron densities, in contradistinction to
the bulk case, resulting in a complicated dependence of the trion
linewidth on $n_{\rm e}$. We note that since there are no
divergences in the exchange terms, replacing $h(\lambda k_1)$ with
$h_s(\lambda k_1)$ in Eqs.\ (\ref{trion_exc}) amounts to a very
small effect (less than $1\%$ in the relevant parameter range).

\vspace*{2.6cm}
\begin{figure}[htbp]
\epsfxsize=0.6\textwidth
\centerline{\epsffile{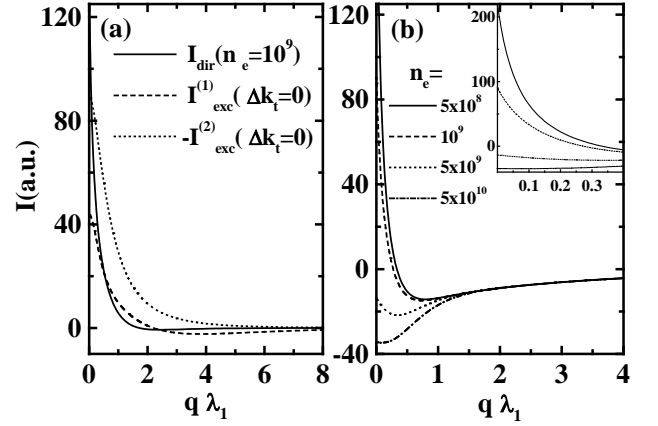}}
\vspace*{-4.3cm}
\caption{Calculated dimensionless matrix elements at $T=5K$ for
the case of $\Delta {\bf k}_{\rm t}=0$, vs.~transferred momentum:
(a) Direct ($n_{\rm e}=10^9 {\rm cm}^{-2}$) and exchange terms
(note that since the exchange terms are not affected by the
screening, their dependence on $n_{\rm e}$ is negligible) ; (b)
Total matrix element for various electron densities $n_{\rm e}$.
The inset provides a close-up at small ${\bf q}$.}
\label{matel_fig}
\end{figure}
Figure \ref{matel_fig}a shows the direct and exchange
dimensionless matrix elements as a function of transferred
momentum $q$, for the case $\Delta {\bf k}_{\rm t}=0$, where the
angular dependence of $V_{\rm exc}^{(1)}$, $V_{\rm exc}^{(2)}$
disappears. The effect of the screening on the total dimensionless
matrix element is shown in Fig.\ \ref{matel_fig}b, where various
electron densities are considered, giving rise to a change in the
screening action. As expected, the effect of the screening is
noticeable in the region of small momenta of the direct matrix
element.

\vspace*{3cm}
\begin{figure}[htbp]
\epsfxsize=0.57\textwidth
\centerline{\epsffile{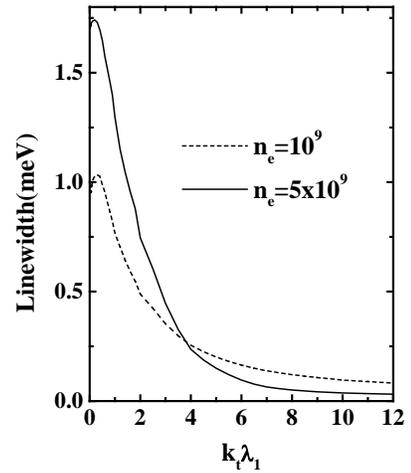}}
\vspace*{-3.6cm}
\caption{Calculated trion linewidth due to elastic scattering vs.
initial trion momentum, at $T=5K$ for two electron densities.}
\label{trion_lw}
\end{figure}
The elastic scattering rate of the trion, $w_{\rm t-e}({\bf
k}_{\rm t} \rightarrow {\bf k}_{\rm t}+{\bf q})$, is calculated by
writing the energy-momentum conservation equation and finding its
solution $k_{\rm e}^*$, similarly to Eqs.\
(\ref{Econs})-(\ref{rate_ex}). Integrating over all final trionic
states, we calculate the trion linewidth due to elastic electron
scattering, $\Gamma_{\rm t-e}(k_{\rm t})$, plotted for two
electron densities in Fig.~\ref{trion_lw}. The linewidth
dependence on $n_{\rm e}$ is quite complicated. First, as we are
considering $T=5K$, the density range for which the linewidth
exhibits a linear dependence on $n_{\rm e}$ is quite small,
similar to the neutral exciton scattering (cf.
Fig.~\ref{fgam_ex}c). Second, the screening potential becomes
larger as the density decreases, thus for very small densities
($n_{\rm e}\lesssim 5\cdot10^8$) decreasing $n_{\rm e}$ results in
a larger trion linewidth. We note that for such low densities, it
is probable that other screening mechanisms become appreciable
(e.g. lattice impurities), thus such a behavior will probably not
be observed experimentally.

\subsection{Electron capture}

We now consider the case in which a heavy-hole exciton captures a
free electron thereby forming a trion. The initial and final
states of the system are given by Eqs.~(\ref{initial}) and
(\ref{tr_st}), respectively. The calculation of the Coulomb
interaction matrix elements between these states essentially
follows the procedure described in the previous section. In this
process there is no exchange term. Instead, the two electrons
which constitute the trion form a singlet spin configuration.
Defining $\Delta {\bf k} \equiv \alpha_{\rm t} {\bf k}_{\rm
t}-\alpha_{\rm x} {\bf k}_{\rm x}$, the calculation yields:
\begin{eqnarray}
&V\!\!&\!\!\!_{\rm cap}(\Delta k)={\cal N}_{\rm cap} \int d^2k_1
h(\lambda_0 k_1) \left\{ \frac{\lambda
\lambda'}{(\lambda_0+\lambda')^2} \times \right. \nonumber \\ &&
\left. \left[ g\left( \tilde{\lambda}' |{\bf k}_1+\Delta {\bf
k}|\right)-g(\tilde{\lambda}' \Delta k) \right] g\left(\lambda
\left|{\bf k}_1+\frac{\Delta {\bf
k}}{\alpha_{\rm x}}\right| \right) + \right. \nonumber \\
&& \left. \frac{\lambda \lambda'}{(\lambda_0+\lambda)^2} \left[
g\left(\tilde{\lambda} |{\bf k}_1+\Delta {\bf
k}|\right)-g(\tilde{\lambda} \Delta k) \right] \times \right.
\\ && \left. g\left( \lambda' \left|{\bf k}_1+\frac{\Delta {\bf
k}}{\alpha_{\rm x}}\right| \right) \right\} \nonumber
\end{eqnarray}
where we have defined the reduced Bohr radii:
$\tilde{\lambda}=\lambda_0 \lambda/(\lambda_0+\lambda)$ and
$\tilde{\lambda}'=\lambda_0 \lambda'/(\lambda_0+\lambda')$, and
the prefactor:
\begin{equation}
{\cal N}_{\rm cap}=\frac{8e^2
\lambda_0^4}{\sqrt{1+\kappa^2}\sqrt{\pi A} L^2\epsilon_0}.
\end{equation}

\begin{figure}[htbp]
\epsfxsize=0.5\textwidth
\centerline{\epsffile{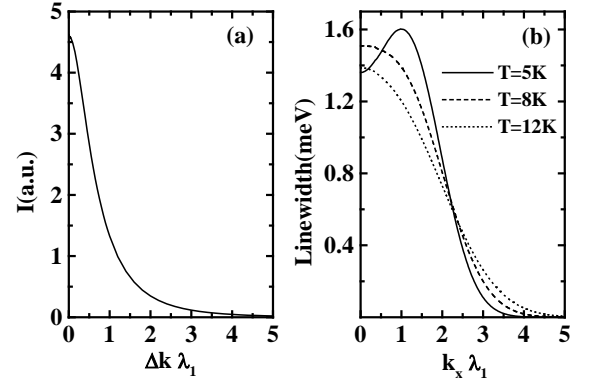}}
\vspace*{-3.4cm}
\caption{(a) Calculated dimensionless matrix element of the
capturing process vs. $\Delta k$ at $T=5K$; (b) $X_{\rm hh}$
calculated linewidth due to capturing scattering for $n_{\rm
e}=10^9 {\rm cm}^{-2}$ at three temperatures.}
\label{fig_extr}
\end{figure}

Fig.\ \ref{fig_extr}a shows the dependence of the dimensionless
matrix element on $\Delta k$. Proceeding as before, we write
conservation of energy:
\begin{equation}
E_{\rm x}(k_{\rm x})+E_{\rm e}({\bf k}_{\rm t}-{\bf k}_{\rm
x})=E_{\rm t}({\rm k}_{\rm t})+E_b^{\rm tr}
\end{equation}
which is satisfied for
\begin{equation}
k_{\rm t}^*=\frac{1}{\beta_t} \left(k_{\rm x} \cos \theta \pm
\sqrt{\beta_t \frac{E_b^{\rm tr} m_{\rm e}}{\hbar^2} -k_{\rm x}^2
\sin^2 \theta} \right)
\end{equation}
where $\theta=\angle({\bf k}_{\rm x},{\bf k}_{\rm t})$. We note
that for the energy conserving solutions, $k_{\rm t}^*$, the
interaction matrix element is constant since $\Delta {\bf k}
(k_{\rm t}^*)=\sqrt{\alpha_x \alpha_t E_b^{\rm tr} m_{\rm
e}/\hbar^2}$. Thus, the dependence of the heavy hole exciton
linewidth on its initial momentum, shown in Fig.\ \ref{fig_extr}b,
comes solely from energy momentum conservation and the Fermi-Dirac
distribution of the scattered electron: $f_{\rm fd}({\bf k}_{\rm
e})= f_{\rm fd}({\bf k}_{\rm t}^*-{\bf k}_{\rm x})$.
\vspace*{1.9cm}

\subsection{Charged exciton dissociation}

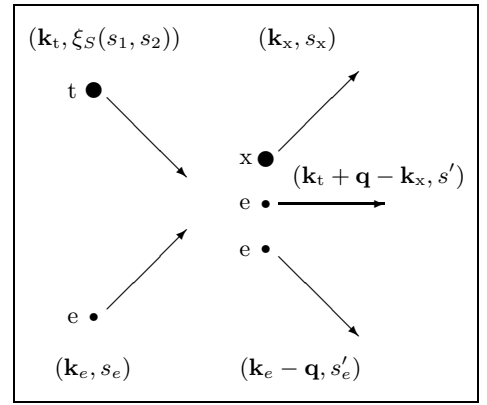
\begin{figure}[htbp]
\vspace*{0.1 cm}
\begin{picture}(160,150)
\put(-5,0){\framebox(175,150)}

\put(15,115){t} \put(25,118){\circle*{6}}
\put(30,115){\vector(1,-1){30}}

\put(0,135){$({\bf k}_{\rm t},\xi_S(s_1,s_2))$}

\put(15,30){e} \put(25,32){\circle*{3}}
\put(30,35){\vector(1,1){30}}

\put(10,10){$({\bf k}_e,s_e)$}

\put(80,90){x} \put(90,92){\circle*{6}}
\put(95,95){\vector(1,1){30}}

\put(87,135){$({\bf k}_{\rm x},s_{\rm x})$}

\put(80,73){e} \put(90,75){\circle*{3}}
\put(95,75){\vector(1,0){40}}

\put(100,83){$({\bf k}_{\rm t}+{\bf q}-{\bf k}_{\rm x},s')$}

\put(80,55){e} \put(90,58){\circle*{3}}
\put(95,55){\vector(1,-1){30}}

\put(80,10){$({\bf k}_e-{\bf q},s'_e)$}

\end{picture}

\caption{Electron-trion inelastic scattering scheme}
\label{trex_scat}
\end{figure}

Lastly, we consider the case in which the scattered trion
dissociates into a neutral exciton and a free electron. This
scattering process is shown schematically in Fig.\
\ref{trex_scat}. The initial state of the system is given by
Eq.~(\ref{initr}) while the final state is that of three free
particles, symbolically written as $|\bf{ k}_{\rm x} ; {\bf
k}_{\rm e} ; {\bf k}_{\rm t}+{\bf q}-{\bf k}_{\rm x} \rangle$,
where ${\bf q}$ is the momentum transferred from the free electron
to the trion, and ${\bf k}_{\rm x}$ is the neutral exciton CM
momentum.

Proceeding as before, we calculate the interaction matrix elements
between the initial and final states of the system. To this end we
denote the final spin state of the three electrons as $|s_{\rm
x},s',s'_e \rangle$ (see Fig.\ \ref{trex_scat}). Assuming the
interaction is spin independent, this final state will be one of
the two states: $(s,m)=(\frac{1}{2},\pm \frac{1}{2})$, where the
state with $m=\frac{1}{2}$ ($m=-\frac{1}{2}$) corresponds to the
initial electron spin $s_e$ being up (down). Actually, we need not
concern ourselves with the exact form of the final electron spin
state, since the projection of the initial spin state (which is a
tensor product of a singlet state with the additional free
electron spin) on a generic  final spin state will extract the
relevant contributions. The proper (anti)symmetrization of this
final spin state is performed inherently by the anticommutation
relations of the electron operators, as in Eq.\ (\ref{delspin}).
As in the elastic scattering case, we find one direct and two
exchange contributions. Denoting $\Delta {\bf k}_{\rm t}={\bf
k}_{\rm e}-\alpha_{\rm t} {\bf k}_{\rm t}$, and $\Delta {\bf
k}_{\rm x}={\bf k}_{\rm e}-\alpha_{\rm x} {\bf k}_{\rm x}$ as
before, these terms are expressed as:
\begin{widetext}
\begin{eqnarray}
V_{\rm dir}(\Delta {\bf k},{\bf q})&=& {\cal N}_{\rm dis}
h_s(\lambda_0 q)\left( \frac{\lambda \lambda'}
{(\lambda_0+\lambda)^2} \left\{ g\left(\lambda' \frac{\Delta
k}{\alpha_{\rm x}}\right) g(\tilde{\lambda}\Delta k) + g
\left(\lambda' \left|\frac{\Delta {\bf k}} {\alpha_{\rm x}}+{\bf
q}\right| \right)  \left[ g(\tilde{\lambda}| \Delta {\bf k}+{\bf
q}|) - g(\tilde{\lambda} \Delta k) \right] \right\} + \right.
\nonumber \\
&& \left. \frac{\lambda \lambda'}{(\lambda_0+\lambda')^2} \left\{
{\renewcommand{\arraystretch}{0.5}
\begin{array}{l}
\lambda \leftrightarrow \lambda' \\
\tilde{\lambda} \rightarrow \tilde{\lambda}'
\end{array}} \right\} \right)
\label{trex_dir}
\end{eqnarray}
and
\begin{subequations}
\label{trex_exc}
\begin{eqnarray}
V^{(1)}_{\rm exc}({\bf q},\Delta{\bf k}_{\rm t},\Delta{\bf k}_{\rm
x})&=& -\frac{\lambda \lambda'}{\pi}{\cal N}_{\rm dis} \int d^2
k_1 h(\lambda_0 k_1) g(\lambda_0 |{\bf k}_1-\Delta {\bf k}_{\rm
x}|) \left( \left\{ \left[ g\left(\lambda \left| {\bf
k}_1-\frac{\Delta {\bf k}}{\alpha_{\rm x}}-{\bf q}\right|\right) -
g\left(\lambda \left| \frac{\Delta {\bf k}}{\alpha_{\rm x}}+{\bf
q}\right|\right) \right] \times  \right. \right. \nonumber \\
&& \left. \left. g(\lambda'| {\bf q}-\Delta {\bf k}_{\rm t}|) +
g(\lambda |{\bf k}_1-\Delta {\bf k}_{\rm t}+{\bf q}|)
g\left(\lambda' \left| \frac{\Delta {\bf k}}{\alpha_{\rm x}}+{\bf
q} \right| \right) \right\}+\{ \lambda \leftrightarrow \lambda' \}
\right) \\
V^{(2)}_{\rm exc}({\bf q},\Delta{\bf k}_{\rm t},\Delta{\bf k}_{\rm
x})&=& -\frac{1}{2}{\cal N}_{\rm dis} h_s\left(\lambda_0 \left|
\frac{\Delta {\bf k}_{\rm x}}{\alpha_{\rm x}}-\frac{\Delta {\bf
k}_{\rm t}}{\alpha_{\rm t}}+{\bf q}\right| \right) \left(
\frac{\lambda \lambda'}{(\lambda_0+\lambda)^2} \left\{ g(\lambda'
|\Delta {\bf k}_{\rm t}-{\bf q}|) g\left(\tilde{\lambda}\left|
\frac{\Delta {\bf k}_{\rm t}-\beta_{\rm x}\Delta {\bf k}_{\rm
x}}{\alpha_{\rm
x}}-{\bf q} \right| \right) + \right. \right. \nonumber \\
&& \left. \left. g(\tilde{\lambda} \Delta k) \left[
g\left(\lambda' \frac{\Delta k}{\alpha_{\rm x}}\right) -
g(\lambda'| \Delta {\bf k}_{\rm t}-{\bf q}|) \right] \right\} +
\frac{\lambda \lambda'}{(\lambda_0+\lambda')^2} \left\{
{\renewcommand{\arraystretch}{0.5}
\begin{array}{l}
\lambda \leftrightarrow \lambda' \\
\tilde{\lambda} \rightarrow \tilde{\lambda}'
\end{array}} \right\} \right),
\label{trex_exc2}
\end{eqnarray}
\end{subequations}
\end{widetext}
where $\tilde{\lambda}$, $\tilde{\lambda}'$ and $\Delta k$ were
defined in the previous subsection and the prefactor multiplying
the dimensionless quantities in Eqs.\
(\ref{trex_dir})-(\ref{trex_exc}), is given by
\begin{equation}
{\cal N}_{\rm dis}=\frac{16 \sqrt{2} e^2 \lambda_0^4
}{\sqrt{1+\kappa^2}L^2 \epsilon_0}
\left(\frac{2\pi}{A}\right)^{3/2}.
\end{equation}

The direct term (\ref{trex_dir}) and the 2nd exchange term
(\ref{trex_exc2}) diverge at $q\rightarrow 0$ and at $(\Delta {\bf
k}_{\rm x}/\alpha_{\rm x}-\Delta {\bf k}_{\rm t}/\alpha_{\rm
t}+{\bf q}) \rightarrow 0$, respectively, implying the need of
using a screened Coulomb potential, as was done in the trion
elastic scattering. Note that the direct term (\ref{trex_dir})
depends only on $k_{\rm t}, k_{\rm x}, q$, and on the angles
$\gamma=\angle ({\bf k}_{\rm t}, {\bf q})$ and $\nu=\angle({\bf
k}_{\rm t},{\bf k}_{\rm x})$, whereas the exchange terms depend
also on the initial electron momentum $k_{\rm e}$ and on the angle
$\delta =\angle ({\bf k}_{\rm e},{\bf q})$. In practice, we
express everything in terms of $q, \Delta k_{\rm t}, \Delta k_{\rm
x}$, and the angles: $\theta_1=\angle(\Delta {\bf k}_{\rm
t},\Delta {\bf k}_{\rm x})$, $\theta_2=\angle(\Delta {\bf k}_{\rm
x},{\bf q})$, and $\theta_3=\angle(\Delta {\bf k}_{\rm t},{\bf
q})$, given by:
\begin{subequations}
\begin{eqnarray}
\cos \theta_1\!\!\!&=\!\!\!&\frac{1}{\Delta k_{\rm x} \Delta
k_{\rm t}} \left(k_{\rm e}^2-\alpha_{\rm x}k_{\rm x} k_{\rm e}
\cos(\gamma-\delta-\nu) - \right. \nonumber \\
&& \left. \alpha_{\rm t} k_{\rm t} k_{\rm e} \cos(\gamma-\delta)+
\alpha_{\rm x} \alpha_{\rm t}k_{\rm x}k_{\rm t}
\cos \nu \right) \\
\cos \theta_2\!\!\!&=\!\!\!&\frac{k_{\rm e} \cos \delta
-\alpha_{\rm x}k_{\rm x} \cos (\gamma-\nu)}{\Delta k_{\rm x}}  \\
\cos \theta_3\!\!\!&=\!\!\!&\frac{k_{\rm e} \cos \delta
-\alpha_{\rm t} k_{\rm t} \cos \gamma}{\Delta k_{\rm t}}.
\end{eqnarray}
\end{subequations}

The calculation of the scattering rate is performed as before,
where conservation of energy now reads:
\begin{eqnarray}
E_{\rm t}(k_{\rm t})+E_{\rm e}(k_{\rm e})&=&E_b^{\rm tr}+E_{\rm
x}(k_{\rm x})+E_{\rm e}({\bf k}_{\rm t}+{\bf q}-{\bf k}_{\rm x})+
\nonumber \\
&& E_{\rm e}({\bf k}_{\rm e}-{\bf q}).
\label{cons_trex}
\end{eqnarray}
Equation (\ref{cons_trex}) is satisfied for
\begin{eqnarray}
k_{\rm e}^*&=&\frac{1}{q \cos \delta} \left\{
\frac{m_e}{\hbar^2}E_b^{\rm tr}+q^2+\frac{1}{2\beta_{\rm t}}
\left(\beta_{\rm t}{\bf k}_{\rm t}-{\bf k}_{\rm x}\right)^2 +
\right. \nonumber \\
&& \left. q \left[ k_{\rm t}\cos \gamma -k_{\rm x} \cos
(\gamma-\nu)\right] \right\}.
\end{eqnarray}
\vspace*{1.2cm}
\begin{figure}[htbp]
\epsfxsize=0.37\textwidth
\centerline{\epsffile{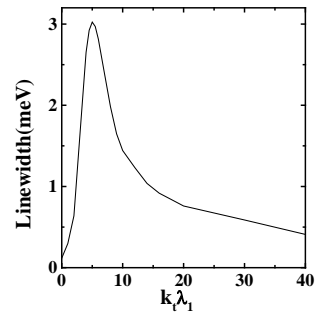}}
\vspace*{-2.6cm}
\caption{Calculated trion linewidth due to inelastic scattering with
electrons vs. its initial momentum, for $n_{\rm e}=10^9 {\rm
cm}^{-2}$.}
\label{lw_trex}
\end{figure}

Using Fermi's golden rule as in Eq.\ (\ref{xrate_ion}), we find
the scattering rate $w_{\rm t-e}({\bf k}_{\rm t} \rightarrow {\bf
k}_{\rm t}+{\bf q})$. Integrating over all final trionic states,
we find the trion linewidth due to inelastic scattering as a
function of its initial momentum ${\bf k}_{\rm t}$, $\Gamma_{{\rm
t-e}}^{\rm dis}({\bf k}_{\rm t})$, shown in Fig.\ \ref{lw_trex}
for $n_{\rm e}=10^9 {\rm cm}^{-2}$.

\section{Excitons and trions in QW's containing a 2DEG}
\label{bare-ex}
\setcounter{equation}{0}

\begin{figure}[htbp]
\epsfxsize=0.45\textwidth
\vspace*{2cm}
\centerline{\epsffile{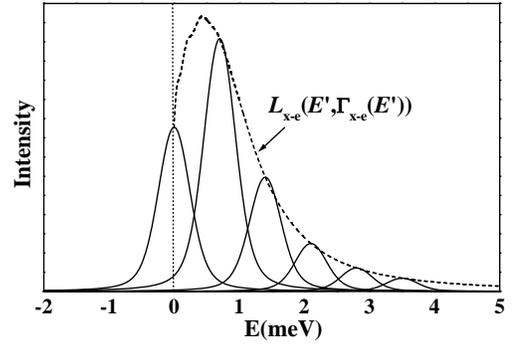}}
\vspace{-3cm}
\caption{A schematic picture of the convolution resulting from
electron scattering. The solid lines represent various exciton
initial line shapes ${\cal I}_0(E-E')$. The contributions of
states with $k_{\rm x}>0$ to the $k_{\rm x}=0$ state are weighted
by the value of the Lorentzian ${\cal L}_{\rm x-e} \left
(E',\Gamma_{\rm x-e}(E') \right)$, given by the dashed line.}
\label{conv}
\end{figure}

In this section we apply the linewidth calculations for the
various scattering processes, presented in the previous sections,
in order to calculate the line shapes of neutral excitons and
trions. These line shapes can be directly compared with measured
reflection spectra, taken from MTQW structures containing a
variable density 2DEG. Without electrons in the QW and for $T
\lesssim 100K$, the two dominant scattering processes are due to
acoustic phonons and static disorder. We note that the radiative
lifetime of a free QW exciton is $\tau_{\rm rad} \approx 25 {\rm
ps}$, resulting in a radiative homogeneous broadening of
$\gamma_{\rm rad} \approx 26 \mu {\rm eV}$. The exciton - acoustic
phonon interaction leads to a homogeneous broadening whose
Lorentzian line shape, ${\cal L}_0$, is characterized by
$\gamma_{\rm phon}=0.2{\rm meV}$ for the $X_{\rm hh}$ and
$\gamma_{\rm phon}=0.28{\rm meV}$ for the $X_{\rm lh}$ at $T=80K$,
using the parameters of GaAs/AlAs QW's \cite{Rap_lw}. The 1S
exciton energy, $E_{\rm 1S}$, has a Gaussian distribution
(characterized by $\gamma_{\rm dis}$) due to the roughness of the
QW surface. Convolving these two distributions results in the
familiar Voigt function \cite{Line_sh}
\begin{equation}
{\cal I}_0(E) = \frac{\gamma_{\rm phon}}{(2 \pi)^{3/2} \gamma_{\rm
dis}} \int_{-\infty}^{\infty} \frac{{\rm e}^{-E_0^2/2 \gamma_{\rm
dis}^2}}{(E-E_0)^2+(\gamma_{\rm phon}/2)^2} dE_0\ , \label{I0}
\end{equation}
where we take $E_{\rm 1S}=0$. When a low density 2DEG is present,
${\cal I}_0(E)$ is convolved with a Lorentzian lineshape that is
associated with the homogeneous broadening that is due to the
electron scattering:
\begin{equation}
{\cal I}(E)= \int_0^{\infty} {\cal I}_0(E-E') {\cal L}_{\rm x-e}
\left (E',\Gamma_{\rm x-e}(E') \right) dE' \ , \label{ls}
\end{equation}
where $\Gamma_{\rm x-e}(E')$ is given by Eq.\ (\ref{Gam_ex}), with
the exciton energy, $E'$, corresponding to $k_{\rm x}$. A
pictorial description of the convolution in Eq.~(\ref{ls}) is
given in Fig.~\ref{conv}, for both elastic and dissociating
exciton-electron scattering processes (the trion-electron
scattering exhibits a similar behavior). It is seen that the
electron scattering admixes excitons having $k_{\rm x}>0$ with the
$k_{\rm x}=0$ state. The degree of admixture is determined by the
value of the Lorentzian that is due to electron scattering, given
in the figure by the dashed line. Note that since $\Gamma_{\rm
x-e}(E')$ is a decreasing function of $E'$, the Lorentzian peak is
shifted to higher exciton energies. (This is observed
experimentally as an increased shift with increasing $n_e$
\cite{prb}).
In the calculation, the Gaussian width, associated with the
inhomogeneity, was taken as a fitting parameter, yielding the
value $\sigma_{\rm dis}=0.2 {\rm meV}$ for both $X_{\rm hh}$ and
$X_{\rm lh}$ (the same value was taken also for the trion line).

\begin{figure}[htbp]
\epsfxsize=0.4\textwidth
\centerline{\epsffile{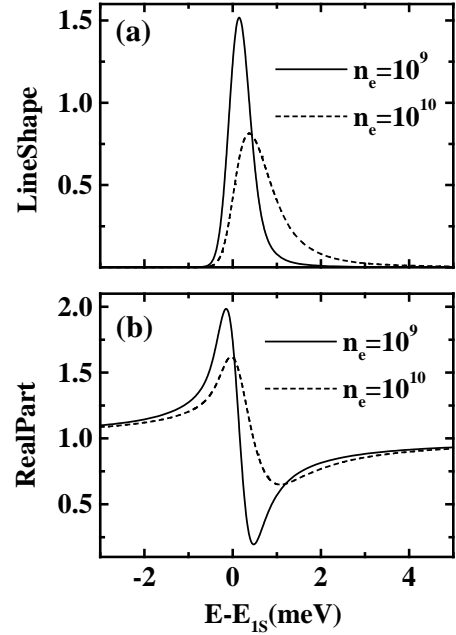}}
\vspace*{-0.8cm}
\caption{(a) Heavy-hole ($M_{\rm x}=0.177m_e$) exciton normalized line
shapes due to scattering with electrons for two electron
densities; (b) The corresponding dielectric function real part.}
\label{line_sh}
\end{figure}

The calculated line shapes are identified with the imaginary part
of the dielectric function in the QW that is related to the
exciton (trion) resonance. Use of the Kramers-Kronig relations
yields the real part of the dielectric function. Examples of
calculated exciton line shapes together with the associated real
part of the dielectric function are depicted in Fig~\ref{line_sh}
for two electron densities.
\vspace*{-0.5cm}

Combining the three exciton line shapes (heavy-hole and light-hole
neutral excitons, and heavy-hole trion), by using the appropriate
oscillator strengths, the reflection spectra of a QW can be
obtained. For high temperature ($T=80K$), where the trion is
dissociated, these calculated spectra and their dependence on
$n_{\rm e}$ were shown to be in good agreement with the
experimental data \cite{prb}.

In the low temperature regime ($2K<T<20K$), the vindication of our
results with experimental measurements is harder to perform. Since
the heavy hole exciton and trion lines are very close in energy,
and the trion oscillator strength is much lower than that of the
exciton, it is practically impossible to discriminate between them
by reflection measurements. PL experiments were performed in MTQW
with 2DEG, at low temperatures, facilitating the observation of
the trion line \cite{Amnon}, but there the detailed balance
between the exciton and the trion line intensities is determined
by the 2DEG density, and the system dynamics should be addressed
by solving, e.g., rate equations.

\vspace*{2.8cm}
\begin{figure}[htbp]
\epsfxsize=0.6\textwidth
\centerline{\epsffile{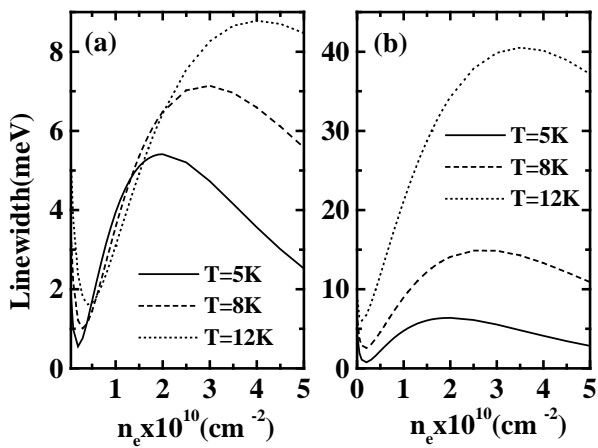}}
\vspace*{-4.2cm}
\caption{Trion's linewidth vs. electron density,
for three temperatures due to (a) elastic scattering ; (b) both
elastic and dissociating scattering.}
\label{amnon}
\end{figure}
In Fig.\ \ref{amnon}a the trion linewidth due to elastic electron
scattering is shown as a function of $n_{\rm e}$ for three
temperatures. The combined effect of both elastic and dissociating
electron scattering on the trion linewidth in shown in Fig.\
\ref{amnon}b. Qualitatively, the trion line observed in
\cite{Amnon} indeed broadens considerably at much lower $n_{\rm
e}$ than the exciton line, as indicated by our results. The
electron densities stated in \cite{Amnon} are probably an
overestimate, as implied by the fact that the neutral exciton
linewidth remains unchanged up to $n_{\rm e}=3 \cdot 10^{10} {\rm
cm}^{-2}$, in clear contradiction to both our calculations and
other experimental data (e.g., \cite{prb}). It should be noted
that calibrating the photoexcitation intensity $I_L$ to $n_{\rm
e}$ in MTQWs is not an easy task. Even if one assumes a linear
dependence of $n_{\rm e}$ on $I_L$ (which is valid for a limited
range), the slope depends on the temperature, since the holes
tunneling times decrease exponentially with $T$, so a different
calibration should be performed not only for different
experiments, but also for each temperature. Finally, the strong
enhancement of the dissociating scattering with increasing
temperature, evident in Fig.\ \ref{amnon}b, implies that the trion
line should practically vanish for $T>10K$ at very low electron
densities.

\section{summary and conclusions}
\label{conc}

In this paper we presented a theoretical study of the various
exciton-electron scattering processes that take place in a QW
which contains a low density 2DEG. We have demonstrated that even
for small densities, of the order of $n_{\rm e} \approx 5 \cdot
10^9 {\rm cm}^{-2}$ in the case of a GaAs/AlAs QW,
exciton-electron scattering is much more efficient than both
exciton-phonon and exciton-exciton scattering, thus playing a
central role in the system dynamics. The dependence of the exciton
linewidth on $n_{\rm e}$ was shown to be drastically different for
the cases of a classical ($T=80K$) and a degenerate ($T=5K$) 2DEG.
In general, the trion-electron scattering poses a much harder
problem, one of the major difficulties being the evaluation of the
screening effects. We showed that the screening model has a
crucial effect on the trion-electron scattering matrix elements,
and hence on the resulting trion linewidth. This is particularly
true for very low densities ($n_{\rm e} \lesssim 5 \cdot 10^8 {\rm
cm}^{-2}$), where the 2DEG screening is less effective, making
other possible mechanisms more dominant. Using the calculated
scattering rates we showed that dissociating scattering can
produce a sizable broadening of the excitonic lines, in particular
for the trion, whose binding energy is considerably smaller than
that of the neutral exciton.

A simple method to produce excitonic line shapes from the
calculated linewidths was devised, allowing our results to be
easily compared with experimental reflection spectra. A reliable
source of experimental data is achievable from reflection
measurements taken from a MTQW structure, embedded in a
microcavity (MC). In the limit of large MC mode energy, the
exciton linewidths are restored, since their coupling with the MC
mode is barely noticeable. A major advantage of MC experiments is
that the trion line is easily resolved by reflection
\cite{Rap2000}, allowing a direct comparison with theory.

The calculated interaction matrix elements can be readily used to
evaluate polariton-electron scattering rates in MCs with 2DEG. In
a recent experiment we have demonstrated polariton final-state
stimulation, assisted by these scattering mechanisms, which was
hitherto observed using polariton-polariton scattering mechanism
\cite{stim_baum,stim_exp}. Incorporation of the various scattering
processes which were considered in this paper, in a consistent
explanation for these nonlinear effects, will be the topic of a
forthcoming publication.

\acknowledgments

This work was supported by the United States - Israel Binational
Science Foundation (BSF), Jerusalem, Israel, by the Fund for
Promotion of Research at the Technion, and by the Technion VPR
Fund - Jewish Communities of Germany Research Fund. G.R.
gratefully acknowledges the financial help from the Technion.


\section*{Appendix A}

\renewcommand{\theequation}{A-\arabic{equation}}
\setcounter{equation}{0}

In this appendix we provide the details of the direct term
calculation. Separating the coordinates in the QW plane ($x-y$)
from the perpendicular coordinate ($z$), and denoting the electron
and hole in-plane momenta by ${\bf k}_1$ and ${\bf k}_2$,
respectively, we write the in-plane Fourier transform of the
exciton wave function $\Phi[({\bf r}_{e\|},z_e),({\bf r}_{h\|},
z_h)]$
\begin{eqnarray}
\Phi_{{\bf k}_1,{\bf k}_2}(z_e,z_h)&=&\frac{1}{A} \int d^2 r_{e\|}
d^2 r_{h\|} \Phi[({\bf r}_{e\|},z_e),({\bf r}_{h\|},z_h)] \times
\nonumber \\
&& {\rm e}^{-{\rm i}({\bf k}_1 \cdot {\bf r}_{e\|} + {\bf k}_2
\cdot {\bf r}_{h\|})},
\end{eqnarray}
where $A$ denotes the QW surface area. Transforming to
center-of-mass (CM) and relative coordinates in the QW plane
(${\bf R}_{\|}= \alpha {\bf r}_{e\|} +\beta {\bf r}_{h\|}, {\bf
r}_{\|}={\bf r}_{e\|}-{\bf r}_{h\|}$, where $\alpha=m_{\rm
e}/M_{\rm x}$, $\beta=1-\alpha$, and $m_{\rm e}$, $M_{\rm x}$ are
the electron and exciton in-plane effective masses, respectively)
we can decompose the exciton wave function into an envelope
function, $\phi (r_{\|},z_e,z_h)$, and a free motion part related
to the in-plane CM coordinate. Denoting ${\bf k}_{\rm x}$ as the
in-plane CM momentum we find
\begin{eqnarray}
\Phi_{{\bf k}_1,{\bf k}_2}(z_e,z_h) &=& \frac{1}{A} \int d^2
R_{\|} d^2 r_{\|} \phi (r_{\|},z_e,z_h) {\rm e}^{{\rm i} {\bf
k}_{\rm x} \cdot {\bf R}_{\|}} \times \nonumber \\
&& {\rm e}^{-{\rm i} \left[ {\bf R}_{\|} \cdot ({\bf k}_1 + {\bf
k}_2)+{\bf r}_{\|}
\cdot (\beta {\bf k}_1 -\alpha {\bf k}_2) \right]} \nonumber \\
&=& \delta({\bf k}_{\rm x}- {\bf k}_1-{\bf k}_2) \phi_{{\bf k}_1
-\alpha {\bf k}_{\rm x}} (z_e,z_h)
\end{eqnarray}

We use the simplest exciton wave function for the exciton ground
state: $\phi (r_{\|},z_e,z_h)={\cal N} \chi_e(z_e) \chi_h(z_h)
{\rm e}^{-r_{\|}/\lambda}$, where ${\cal N}$ is a normalization
factor and $\lambda$ is a variational parameter associated with
the exciton Bohr radius in the QW, which is fixed by maximizing
the binding energy of the exciton \cite{Bastard}. We note that the
use of a wave function separable in $z$ and $r_{\|}$, facilitates
considerably the calculation of the scattering matrix elements,
although it is strictly justifiable only for narrow well
structures. Assuming perfect confinement of electrons and holes in
the QW, and taking the $z$ axis origin in the center of the QW,
the confinement functions are:
\begin{equation}
\chi_e(z_e)=\chi_h(z_h)= \left\{
\begin{array}{ll}
\cos (\pi z/L) \ & \left|z\right| \leq L/2 \\
0 \ & \left| z \right| >L/2
\end{array}
\right.
\label{chi}
\end{equation}
This approximation is quite accurate for almost all practical
thicknesses $L$. Calculating the binding energy of the 1S exciton
in a GaAs QW of width $L=200$\AA, we find $E_b=6.73 {\rm meV},
\lambda=150$\AA ($E_b=7.49 {\rm meV}, \lambda=133$\AA) for the
heavy-hole (light-hole) exciton. These results, obtained with our
single parameter trial function, are quite close to those obtained
in \cite{GruBim}, which took into account finite QW barriers, and
anisotropic masses.

Combining the first terms from Eqs. (\ref{matel}), the direct term
reads
\begin{eqnarray}
V_{\rm dir}(q) &=& \frac{32 \lambda^2 e^2}{\epsilon_0 AL^3 q}
\int_{-L/2}^{L/2} dz_e dz_h dz_c \int d^2 k_1 {\rm e}^{-q
|z_e-z_c|} \times \nonumber \\
&&\left[  g(\lambda| \beta {\bf q}-{\bf k}_1|) - g(\lambda|\alpha
{\bf q}+{\bf k}_1|) \right] g(\lambda k_1)
\times \nonumber \\
&& \cos^2(\pi z_e/L) \cos^2(\pi z_h/L) \cos^2(\pi z_c/L),
\end{eqnarray}
where $g(\lambda q)$ was defined in Eq. (\ref{gh}). We note that
the electron wave functions in the QW plane are absent, as they
contribute, together with the in-plane center of mass part of the
exciton wave function, a fixed phase factor which is unimportant
for our purposes. The momentum integral is a simple convolution
and can be readily evaluated
\begin{eqnarray}
&& \int d^2 k_1 \left[ g(\lambda| \beta {\bf q}-{\bf k}_1|)
-g(\lambda|\alpha{\bf q}+{\bf k}_1|) \right] g(\lambda k_1) \nonumber \\
&=& \frac{1}{4 \pi^2} \int d^2 \rho \tilde{g}^2(\rho/\lambda)
\left[ \exp ({\rm i} \mbox{\boldmath $\rho$} \cdot \beta {\bf q})
- \exp ({\rm i} \mbox{\boldmath $\rho$} \cdot \alpha {\bf q})
\right] \nonumber \\
&=& \frac{\pi}{2 \lambda^2} \left[ g(\lambda
\beta q/2) - g(\lambda \alpha q/2) \right]
\end{eqnarray}
where
\begin{equation}
\tilde{g}(\rho/\lambda) = \frac{2 \pi}{\lambda^2} {\rm e}^{-\rho/
\lambda}
\end{equation}
is the exciton spatial wave function in the QW plane. In order to
evaluate the $z$ integrals, we change to the coordinates
\[
z=z_e-z_c \ \ ; \ \ z'=\frac{1}{2} (z_e+z_c)
\]
which implies the following change in the integration limits:
\begin{eqnarray*}
\int_{-L/2}^{L/2} dz_e \int_{-L/2}^{L/2} dz_c&=&\int_{-L/2}^{0}
dz' \int_{-(L+2 z')}^{L+2 z'} dz \\ &+& \int_{0}^{L/2} dz'
\int_{-(L-2z')}^{L-2z'} dz,
\end{eqnarray*}
and performing the integrations results in the direct term, given
by Eq. (\ref{Vdir})

\section*{Appendix B}

\renewcommand{\theequation}{B-\arabic{equation}}
\setcounter{equation}{0}

In this Appendix we present an analytical calculation of the
angular part of the exchange integral. This calculation, while
quite complicated, is worth the trouble for two reasons. The first
is, obviously, a major reduction in the time consumed by the
numerical computation, making it feasible. As a subsidiary
benefit, the partially analytical calculation removes some
convergence problems encountered in the previously used numerical
algorithm.

Using dimensionless momenta (e.g., by writing $q$ we mean $\lambda
q$) the exchange integral reads
\begin{widetext}
\begin{eqnarray}
I_{\rm exc}(\Delta k,q,\theta) &=& -4 \pi \int dk_1 k_1 h(k_1) d
\phi  \left\{ \left[ 1+({\bf k}_1- \Delta {\bf k}+{\bf q})^2
\right]^{-3/2}-\left[ 1+({\bf q}- \Delta {\bf k})^2 \right]^{-3/2}
\right\} \times \nonumber \\
&& \left[1+({\bf k}_1- \Delta {\bf k}+\alpha {\bf q})^2
\right]^{-3/2}  \equiv -4 \pi \int dk_1 k_1 h(k_1) \left\{ (I) +
(II) \right\} ,
 \label{genI}
\end{eqnarray}
\end{widetext}
where $\theta$ is the angle between ${\bf q}$ and $\Delta {\bf k}$
and $\phi$ is the angle between ${\bf k}_1$ and $\Delta {\bf k}$.
We shall work out separately the two parts of Eq.\ (\ref{genI}).

The first part is given by
\begin{eqnarray}
(I)&=& \int_0^{2 \pi} d \phi \left[ ad+(bd+ae)\cos \phi
- \right. \nonumber \\
&& \left. c(d+a/\alpha) \sin \phi - c(e+b/\alpha) \sin \phi \cos
\phi + \right. \nonumber \\
&& \left. be \cos^2 \phi +c^2/\alpha \sin^2 \phi \right]^{-3/2},
\end{eqnarray}
where we have defined
\begin{eqnarray}
a &=& 1+k_1^2+\Delta k^2+\alpha^2 q^2 -2 \alpha q \Delta k \cos
\theta \nonumber \\
b &=& 2 k_1 (\alpha q \cos \theta -\Delta k) \nonumber \\
c &=& 2 \alpha q k_1 \sin \theta \\ \label{abcd}
d &=& 1+k_1^2+\Delta k^2+q^2 -2 q \Delta k \cos \theta \nonumber \\
e &=& 2 k_1 (q \cos \theta -\Delta k) \nonumber.
\end{eqnarray}
Transforming $\phi \rightarrow \phi'=\phi-\pi$ and then to $z=\tan
( \phi/2)$ we have
\begin{eqnarray*}
(I) &=& 2 \int_{-\infty}^{\infty} dz (1+z^2)^2 \left[
(a-b)(d-e)+\right. \\ && \left. 2c(d-e+a/\alpha-b/\alpha)z+
2(ad-be+2c^2/\alpha)z^2+\right. \\
&& \left. 2c(d+e+a/\alpha+b/\alpha)z^3+(a+b)(d+e)z^4
\right]^{-3/2}.
\end{eqnarray*}
Finding the roots of this 4th order polynomial we have
\begin{eqnarray}
(I)&=& 2 \int_{-\infty}^{\infty} dz (1+z^2)^2 \left[(z^2+2Az+B)
\times \right. \nonumber \\
&& \left. (z^2+2Cz+D)(a+b)(d+e) \right]^{-3/2}
\label{I1}
\end{eqnarray}
where
\[
A=\frac{c}{a+b} \ \ ; \ \ B=\frac{a-b}{a+b} \ \ ; \ \
C=\frac{c}{\alpha(d+e)} \ \ ; \ \ D=\frac{d-e}{d+e} \ .
\]
We note for later use that $A,B,C,D$ obey the inequalities
\begin{equation}
B>0 \ \ ; \ \ D>0 \ \ ; \ \ B-A^2>0 \ \ ; \ \
D-C^2>0 \ .
\label{ineq}
\end{equation}
One can reduce the polynomial in Eq.\ (\ref{I1}) to a product of
two quadratic binomial forms, by making use of the following
transformation
\[
z=\frac{p+qy}{1+y}
\]
with $p$, $q$ given by
\[
p,q=\frac{1}{2} \frac{D-B}{A-C} \pm
\sqrt{\frac{1}{4} \left(\frac{D-B}{A-C}
\right)^2+\frac{AD-BC}{A-C}}
\]
(note that we assume $A \neq C$, deferring for later the treatment
in this special case). Noticing that $p$ and $q$ always have
opposite signs, we choose $q<0$ and divide the integral into
\[
\int_{-\infty}^{\infty}dz=\int_{-\infty}^0 dz+\int_0^{\infty} dz
\]
where the limit $z=\infty$ corresponds to $y=-1$ and the limit
$z=-\infty$ corresponds to $y=-1$ with $q$ and $p$ interchanged.
The integral now reads
\begin{widetext}
\[
(I) = {\cal N} \left\{ \int_{-p/q}^{-1} dy \frac{ \left[
y^2(1+q^2)+2y(1+pq)+1+p^2 \right]^2}{ \left[ (y^2+s^2)(y^2+t^2)
\right]^{3/2}} - \frac{1}{(st)^3} \int^{-q/p}_{-1} dy \frac{
\left[ y^2(1+p^2)+2y(1+pq)+1+q^2 \right]^2}{ \left[
(y^2+1/s^2)(y^2+1/t^2) \right]^{3/2}} \right\}
\]
where
\[
s^2=\frac{p^2+2Ap+B}{q^2+2Aq+B} \ \ \ ; \ \ \
t^2=\frac{p^2+2Cp+D}{q^2+2Cq+D},
\]
and
\[
{\cal N} =\frac{2(q-p)}{\left[(q^2+2Aq+B)(q^2+2Cq+D)(a+b)(d+e)
\right]^{3/2}}.
\]
Making the transformation $y=s \tan \psi$ in the first integral
and $y=\frac{1}{s}\tan \psi$ in the second, we find
\begin{eqnarray*}
(I) &=& \frac{{\cal N}}{s^2t^3} \left\{ \int_{\tan^{-1}
(-p/sq)}^{\tan^{-1} (-1/s)} d \psi  \frac{\left[ (qs \sin \psi +p
\cos \psi)^2+(s \sin \psi + \cos \psi)^2 \right]^2}{(1-k^2 \sin^2
\psi)^{3/2}} - \right. \\ && \left. st^3 \int_{\tan^{-1}
(-s)}^{\tan^{-1} (-qs/p)} d \psi \frac{\left[ (p/s \sin \psi +q
\cos \psi)^2+(1/s \sin \psi + \cos \psi)^2
\right]^2}{(1-\tilde{k}^2 \sin^2 \psi)^{3/2}} \right\},
\end{eqnarray*}
where we have defined
\begin{equation}
k^2=1-s^2/t^2 \ \ \ ; \ \ \ \tilde{k}^2=1-t^2/s^2.
\label{k}
\end{equation}
Changing $\psi \longrightarrow \pi/2-\psi$ in the second integral,
and using $\tan^{-1}s+\cot^{-1}s=\pi/2$ we find that the second
integral exactly matches the first except that its upper limit is
larger by $\pi$; thus we are left with
\[
(I)=-\frac{{\cal N}}{s^5} \int_{-\tan^{-1}s}^{\pi-\tan^{-1}s}
\!\!\!\!\! d \psi \frac{\left[ (1+p^2) \sin^2 \psi +s^2(1+q^2)
\cos^2 \psi +s(1+pq) \sin 2\psi \right]^2}{(1-\tilde{k}^2 \sin^2
\psi)^{3/2}}.
\]
This integral can be evaluated using incomplete elliptic integrals
(see p.\ 201 in \cite{gd}). Furthermore, one can exploit the
functional relations (p.\ 911 in \cite{gd})
\begin{eqnarray*}
F(-\phi,k)=-F(\phi) \ \ \ &;& \ \ \ F(n \pi - \phi,k)=2 {\bf K}(k)-F(\phi,k)
\nonumber \\
E(-\phi,k)=-E(\phi) \ \ \ &;& \ \ \ E(n \pi - \phi,k)=2 {\bf
E}(k)-E(\phi,k)
\end{eqnarray*}
where $F(\phi ,k),E(\phi ,k)$ denote incomplete elliptic integrals
of the first and second kind, respectively, and ${\bf K}(k), {\bf
E}(k)$ denote complete elliptic integrals of the first and second
kind, respectively. The final result is
\begin{eqnarray}
(I) &=& -\frac{2{\cal N}}{s(s^2-t^2)^2} \left\{ \left[ (s^2+t^2)
{\bf E}(\tilde{k})-2t^2{\bf K}(\tilde{k})\right]\left[
\frac{1}{t^2}(1+p^2)^2+(1+q^2)^2s^2 \right] - \right. \nonumber \\
&& \left. 2(3+3p^2q^2+4pq+p^2+q^2) \left[ 2s^2 {\bf
E}(\tilde{k})-(s^2+t^2) {\bf K}(\tilde{k}) \right] \right\} \ .
\label{Ires}
\end{eqnarray}
\end{widetext}
For imaginary values of $\tilde{k}$, the result is the same as
Eq.\ (\ref{Ires}) with the interchanges $s \longleftrightarrow t $
and $\tilde{k} \longrightarrow k$, $k$ being defined in (\ref{k}).

We now return to examine two special cases which cannot be treated
using (\ref{Ires}). The first case corresponds to $A=C \ , B \neq
D$. Here we have
\[
p= \pm \infty \ \ \ ; \ \ \ q=-A.
\]
Although $s,t =\infty$, $\tilde{k}$ is still finite, taking the
value
\[
\tilde{k}^2=\frac{D-B}{D-A^2},
\]
thus, enabling the use of (\ref{Ires}) with some modifications.
Remembering the inequalities (\ref{ineq}) we note that the sign of
$\tilde{k}^2$ is determined by the sign of $D-B$. We obtain, for
$D>B$
\begin{eqnarray}
&(\!\!\!&I) = \frac{4 \sqrt{D-A^2}}{\left[ (a+b)(d+e)
\right]^{3/2}
(D-B)^2} \times \label{Ires1} \\
&&\!\!\! \left\{ \left[1+ \frac{(1+A^2)^2}{(B-A^2)(D-A^2)} \right]
\left[ (1+D-A^2){\bf E}(\tilde{k})- \right. \right. \nonumber \\
&&\!\!\! \left. \left. 2 {\bf K}(\tilde{k}) \right] \!\!-
2(3A^2+1) \!\! \left[ 2{\bf E}(\tilde{k})-\!\! \left(
1+\frac{B-A^2}{D-A^2} \right){\bf K}(\tilde{k}) \right] \right\}.
\nonumber
\end{eqnarray}
For $D<B$ the result is the same as Eq.\ (\ref{Ires1}) with the
interchanges $B \longleftrightarrow D $ and $\tilde{k}
\longrightarrow k$.

The second special case corresponds to $A=C \ , B=D$. Here the
integral (\ref{I1}) reduces to
\[
(I)=\frac{2}{[(a+b)(d+e)]^{3/2}} \int_{-\infty}^{\infty} dz
\frac{(1+z^2)^2}{(z^2+2Az+B)^3}.
\]
This integral can be easily evaluated (see p.\ 81 in \cite{gd})
and the result is
\begin{equation}
(I)=\frac{\pi}{4}
\frac{3+4A^2+2B+3B^2}{[(a+b)(d+e)]^{3/2}(B-A^2)^{5/2}}.
\end{equation}

Finally, we evaluate the second part of the exchange integral,
appearing in Eq.\ (\ref{genI}). We have
\[
(II)=\frac{1}{(d-k_1^2)^{3/2}} \int_0^{2 \pi} \frac{d \phi}{(a+b
\cos \phi -c \sin \phi)^{3/2}} ,
\]
where $a,b,c,d$ where defined in (\ref{abcd}). Using the
transformation $\phi=2 \psi +\beta$, where $\tan \beta =-c/b$, we
have
\[
(II)= \frac{2}{(d-k_1^2)^{3/2}}
\int_{-\beta/2}^{\pi- \beta/2} \frac{d \psi}{(a+p-2p\sin^2
\psi)^{3/2}},
\]
where $p=\sqrt{b^2+c^2}$. This integral can be evaluated using the
elliptic integral of second kind, resulting in
\begin{equation}
(II)=\frac{4 \sqrt{a+p}}{(d-k_1^2)^{3/2}(a^2-p^2)} {\bf E} \left(
\frac{2p}{a+p} \right).
\end{equation}

\section*{Appendix C}

\renewcommand{\theequation}{C-\arabic{equation}}
\setcounter{equation}{0}

In this Appendix we present the calculation of the trion binding
energy, based on the trial wave function given in Eq.\
(\ref{phitr}). It is useful to work with the Fourier transformed
trion wave function (\ref{tr_k}), where we have for the two spin
configurations, explicitly
\begin{eqnarray*}
\phi^{\rm tr}_{{\bf k}_1,{\bf k}_2}(z_1,z_2,z_h)&=&{\cal N}_{\rm
tr} \chi(z_1) \chi(z_2) \chi(z_h) \times \nonumber \\
&&\left[ \phi_{{\bf k}_1}\phi'_{{\bf k}_2}\pm \phi'_{{\bf
k}_1}\phi_{{\bf k}_2} \right].
\end{eqnarray*}
The effective mass Hamiltonian is given by
\begin{widetext}
\begin{equation}
H=\sum_{i=1}^{2} \left[ -\frac{\hbar^2}{2m_e} \left(
\nabla^2_{{\bf r}_i}+\partial^2_{z_i} \right) -\frac{e^2}
{\epsilon_0\sqrt{({\bf r}_i-{\bf r}_h)^2+(z_i-z_h)^2}} \right]+
-\frac{\hbar^2}{2 m_e}\left(\sigma \nabla^2_{{\bf r}_h} +\sigma_z
\partial^2_{z_h} \right)+ \frac{e^2}{\epsilon_0\sqrt{({\bf
r}_1-{\bf r}_2)^2+(z_1-z_2)^2}},
\end{equation}
where $\sigma=m_e/m_h^{||}$, $\sigma_z=m_e/m_h^{z}$ are the
electron-hole effective mass ratios in the plane and $z$
directions, respectively. Transforming to CM and relative
coordinates in the QW plane:
\begin{subequations}
\begin{eqnarray}
{\bf R}_{\rm cm}&=&\frac{\sigma({\bf r}_1+{\bf r}_2)+{\bf
r}_h}{1+2 \sigma} \\
{\bf r}_{\rm ih}&=&{\bf r}_i-{\bf r}_h, \ \ i=1,2
\end{eqnarray}
\end{subequations}
and neglecting CM motion which does not contribute to the binding
energy, we have
\begin{equation}
H=\sum_{i=1}^{2} H_0({\bf r}_{\rm ih},z_i,z_h)+H_{12}({\bf r}_{\rm
1h},z_1,{\bf r}_{\rm 2h},z_2),
\end{equation}
where
\begin{subequations}
\begin{eqnarray}
H_0({\bf r}_{\rm ih},z_i,z_h)&=&-(1+\sigma) \nabla^2_{{\bf r}_{\rm
ih}} -\frac{\sigma_z}{2} \partial^2_{z_i} -\frac{2}{\sqrt{r_{\rm
ih}^2+(z_i-z_h)^2}} \\
H_{12}({\bf r}_{\rm 1h},z_1,{\bf r}_{\rm 2h},z_2)&=&-2\sigma
\nabla_{r_{\rm 1h}} \cdot \nabla_{r_{\rm 2h}}+\frac{2}{\sqrt{({\bf
r}_{\rm 1h}-{\bf r}_{\rm 2h})^2+(z_1-z_2)^2}},
\end{eqnarray}
\end{subequations}
and we employed natural units of length and energy, namely, bulk
effective Bohr radius and Rydberg. The expectation value of the
trion energy is given by
\begin{equation}
E=\langle \phi^{\rm tr} | H | \phi^{\rm tr} \rangle=\frac{{\cal
I}_1 \pm 2\kappa{\cal I}_2+{\cal J}_1 \pm {\cal J}_2}{1 \pm
\kappa^2},
\label{Etr}
\end{equation}
where $\kappa$ was defined in Eq.\ (\ref{s}), and the upper
(lower) signs apply to the singlet (triplet) spin configuration.
The various terms appearing in Eq.\ (\ref{Etr}) are calculated in
the in-plane Fourier space, resulting in simple 2D convolutions,
similar to those in appendix A. The results are:
\begin{subequations}
\begin{eqnarray}
{\cal I}_1&=&\left(\frac{2}{L}\right)^2\left[ \langle \phi \chi_1
\chi_h |H_0|\phi \chi_1 \chi_h \rangle +\langle \phi' \chi_1
\chi_h |H_0|\phi' \chi_1 \chi_h \rangle \right]= \left(
\frac{\pi}{L} \right)^2(2+\sigma_z) + \nonumber \\
&&\left[\frac{1}{\lambda^2}(1+\sigma)-\lambda^3\left(\frac{2}{L}
\right)^2 \int_0^{\infty}dq q h(\lambda q)g(\lambda q/2)
\right]+[\lambda \rightarrow \lambda'] \\
{\cal
I}_2&=&\left(\frac{2}{L}\right)^2 \langle \phi \chi_1 \chi_h |H_0|
\phi' \chi_1 \chi_h \rangle = \frac{\kappa}{2}\left(
\frac{\pi}{L}\right)^2 (2+\sigma_z)+
\frac{4(1+\sigma)}{(\lambda+\lambda')^2}-(\lambda+\lambda')
\left(\frac{\tilde{\lambda}\kappa}{L}\right)^2 \int_0^{\infty}dq q
h(\tilde{\lambda}q) g(\tilde{\lambda}q) \\
{\cal J}_1&=&\left(\frac{2}{L}\right)^2 \langle \phi_1 \chi_1
\phi'_2 \chi_2|H_{12}|\phi_1 \chi_1 \phi'_2 \chi_2 \rangle =
2\lambda^3 \left(\frac{2}{L}\right)^2 \int_0^{\infty}dq q
h(2\lambda q) g(\lambda q/2)g(\lambda'q/2) \\
{\cal J}_2&=&\left(\frac{2}{L}\right)^2 \langle \phi_1 \chi_1
\phi'_2 \chi_2 |H_{12}|\phi'_1 \chi_1  \phi_2 \chi_2  \rangle =
2\kappa^2\tilde{\lambda}^3 \left(\frac{2}{L}\right)^2
\int_0^{\infty}dq q h(2\tilde{\lambda}q) g^2(\tilde{\lambda}q).
\end{eqnarray}
\label{IJ}
\end{subequations}
In Eqs.\ (\ref{IJ}), $h(\lambda q),g(\lambda q)$ are the functions
defined in Eq.\ (\ref{gh}), and we have denoted
$\tilde{\lambda}=\lambda \lambda'/(\lambda+\lambda')$. We note
that in the 2D limit ($L \rightarrow 0$), eliminating the kinetic
energy term in the $z$ direction, Eqs.\ (\ref{IJ}) reduce to
\begin{subequations}
\begin{eqnarray}
{\cal I}_1&=&\left(\frac{1}{\lambda^2}+\frac{1}{\lambda^{\prime
2}}\right)
(1+\sigma)-4\left( \frac{1}{\lambda}+\frac{1}{\lambda'} \right) \\
{\cal I}_2&=&\frac{4(1+\sigma)}{(\lambda+\lambda')^2}-\frac{8}
{\lambda+\lambda'} \\
{\cal J}_1&=&\left\{
\begin{array}{ll}
\frac{\lambda \kappa}{(\lambda'-\lambda)^2}\left[ \left(1+
\frac{\lambda^{\prime 2}}{\lambda^2}\right){\bf E}(1-\lambda^2/
\lambda^{\prime 2})-2 {\bf K}(1-\lambda^2/ \lambda^{\prime
2})\right], &
\lambda<\lambda' \\
\frac{\lambda' \kappa}{(\lambda'-\lambda)^2}\left[ \left(1+
\frac{\lambda^2}{\lambda^{\prime 2}}\right){\bf
E}(1-\lambda^{\prime 2}/ \lambda^2)-2 {\bf K}(1-\lambda^{\prime
2}/ \lambda^2)\right], &
\lambda>\lambda' \\
\frac{3 \pi}{4 \lambda}, & \lambda=\lambda'
\end{array}
\right. \\
{\cal J}_2&=&\frac{3 \pi \kappa}{2 (\lambda+\lambda')}
\end{eqnarray}
\end{subequations}
\end{widetext}
where ${\bf K}$, ${\bf E}$ are the complete elliptic integrals of
the first and second kind, respectively. These closed form
expressions coincide with the integral expressions of Sandler and
Proetto \cite{SanPro}, for $\sigma \rightarrow 0$ and vanishing
magnetic field (note a misprint in Eq.\ A6 of this paper).

In order to maximize the trion binding energy, we must first write
the expectation value of the exciton energy:
\begin{eqnarray}
E_0&=&\frac{1}{\lambda_0}(1+\sigma)+\left(\frac{\pi}{L}\right)^2
(1+\sigma_z)-\nonumber \\
&& \frac{4 \lambda^3_0}{L^2} \int_0^{\infty}dq q h(\lambda_0
q)g(\lambda_0 q/2),
\end{eqnarray}
resulting in the 1S exciton Bohr radius and binding energy values
stated in Appendix A. Note that in the 2D limit we have
\begin{equation}
E_0(L\rightarrow 0)=\frac{1+\sigma}{\lambda_0^2}-
\frac{4}{\lambda_0}.
\label{Ex0}
\end{equation}
Minimizing Eq.\ (\ref{Ex0}) in the $\sigma \rightarrow 0$ limit,
results in $\lambda_0=a_{\rm B}/2$, $E_0=-4{\rm Ry}$, as expected.
The trion binding energy with respect to the exciton is related to
the trion and exciton energies through
\begin{equation}
E_b^{\rm tr}=E-E_0-E_{\rm e},
\label{Ebtr}
\end{equation}
where $E_{\rm e}=(\pi/L)^2$ is the confined electron energy, in
our infinite barrier model. We note that the kinetic energy terms
in the $z$ direction cancel out in the expression for the trion
binding energy, so they can be removed from the calculation.
Maximizing Eq.\ (\ref{Ebtr}) with respect to the two variational
parameters $\lambda,\lambda'$, we finally find the results given
in section \ref{trion}, where we used the GaAs bulk values of the
effective Bohr radius ($a_B=100$\AA), and Rydberg (5.7meV).



\begin{thebibliography}{99}
\bibitem{delta} The angle between ${\bf k}_{\rm x}$ and
${\bf k}_{\rm e}$ is thus given by $\gamma \pm \delta$, where it
is noted that since one can make the transformation $\delta
\rightarrow 2 \pi -\delta$, the result of the integration over
$\delta$ is indifferent to the sign choice in $\cos(\gamma \pm
\delta)$.
\end{thebibliography}
\end{document}